\documentclass[5p]{elsarticle}

\usepackage{ifpdf}
\ifpdf\usepackage{epstopdf}\fi
\usepackage{amsmath,amssymb}

\usepackage{lineno,hyperref}
\modulolinenumbers[5]

\journal{Journal of \LaTeX\ Templates}

\biboptions{sort&compress}

\begin{document}

\begin{frontmatter}

\title{On certain aspects of the THERMOS toolkit for modeling experiments}

\author{I.Yu. Vichev\corref{mycorrespondingauthor}}
\cortext[mycorrespondingauthor]{Corresponding author}
\ead{vichevilya@keldysh.ru}

\author{A.D. Solomyannaya, A.S. Grushin, D.A. Kim}
\address{Keldysh Institute of Applied Mathematics RAS, Miusskaya sq.4, 125047 Moscow, Russian Federation}

\begin{abstract}
The THERMOS toolkit has been developed to calculate radiative properties of plasmas. This article contains a brief survey of some of its key features used by calculation of opacities and emissivities and by analysis of specific experiments. The code has recently been upgraded to account for the effect of ionization potential lowering in dense plasmas. The functionality of the code is illustrated for several cases from the 10$^{th}$ NLTE Code Comparison Workshop, in particular, for the experimental spectra of chlorine~\cite{doi:10.1063/1.4965233} and for the measured transmission of a silicon plasma~\cite{PhysRevLett.119.075001}.
\end{abstract}


\end{frontmatter}


\section{Introduction}
Originally, the THERMOS code was developed at KIAM RAS (Keldysh Institute of Applied Mathematics of the Russian Academy of Sciences) by the research team of A.F.~Nikiforov, V.B.~Uvarov and V.G.~Novikov to calculate the properties of equilibrium plasmas in a wide range of temperatures and densities. It is based on the non-relativistic modified Hartree-Fock-Slater model in the average atom approximation and is described in the book ``Quantum Statistical Models of High-Temperature Plasma'' \cite{Nikiforov2005}. Spectral opacity tables, calculated with the THERMOS code, have been extensively used to solve various problems of radiation dominated plasmas; see for example Refs.~\cite{FAIK201447,1742-6596-653-1-012148, doi:10.1063/1.4921334,doi:10.1063/1.4960684}.

The development of new powerful facilities, such as NIF and Z, gave rise to a large amount of experimental data for non-equilibrium plasmas that resulted in revision of many existing models. Additional strong motivation came also from the development of extreme ultraviolet light sources, where simulations require high accuracy of line positioning and new methods for computing the spectral tables. These problems gave a major impetus for further development of the THERMOS toolkit.

A number of codes, developed earlier by our research team for calculating the radiative properties of plasmas, have recently been consolidated into the THERMOS toolkit software package. In result, it has become a modern instrument that allows calculation of the emission and absorption spectra in a wide range of temperatures and densities, both in the approximation of local thermodynamic equilibrium (LTE), and for the non-equilibrium (NLTE) plasmas with an arbitrary radiation field. The newly implemented models take into account the density effects by allowing for the ionization potential depression (IPD) according to the Stewart-Pyatt \cite{stewart_lowering_1966} and/or the Ecker-Kr{\"o}ll~\cite{doi:10.1063/1.1724509} formulae, and for the ion broadening of spectral lines in the two-level approximation~\cite{Rozsnyai1977a}. Also, a new code module has been added where the equations of radiation transfer and level kinetics are solved self-consistently in a steady-state approximation across a multi-layer slab sample.

The THERMOS toolkit is a regular participant at the NLTE Code Comparison Workshops~\cite{NLTE_site}, and its calculation results demonstrate a fair agreement with the other well-known codes from all over the world. At the 10$^{th}$ NLTE Workshop (2017, San Diego, CA, USA), the experimentally obtained emission and transmission spectra for steady-state cases of the aluminum, chlorine, and silicon plasmas were proposed as benchmark problems for simulation. Some of the simulation results, obtained with the THERMOS toolkit for this workshop, were recently presented at the RPHDM18~\cite{RPHDM18} conference and are described in some detail below.

\section{THERMOS toolkit overview}
\label{subSection_Toolkit_structure}

\begin{figure}[!htb]
	\centering
	\includegraphics[width=0.9\linewidth]{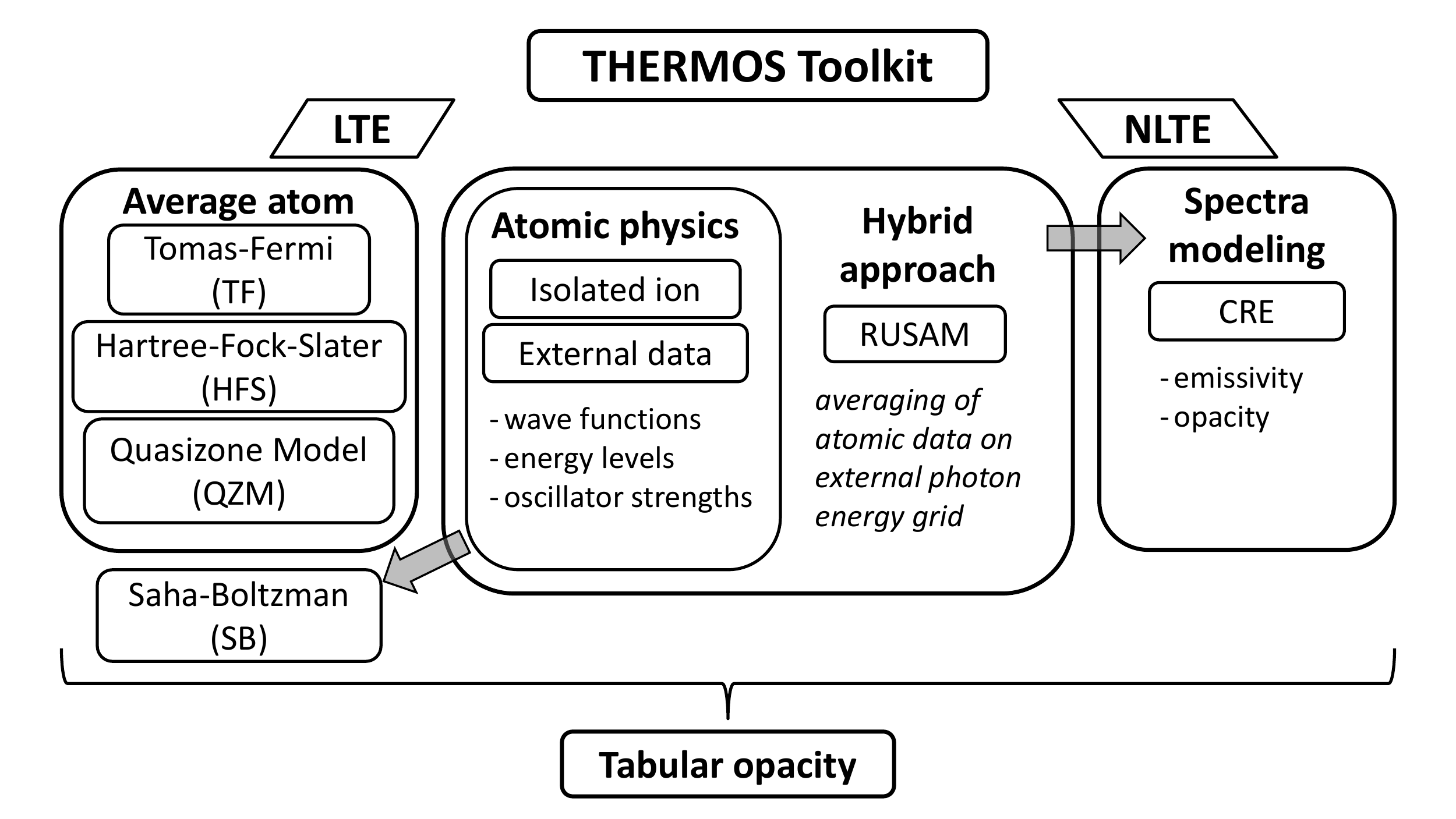}
	\caption{THERMOS toolkit overview.}
	\label{THERMOS_Toolkit_structure}
\end{figure}

The THERMOS toolkit includes codes for numerical simulation of the LTE and non-LTE plasmas, see Fig.~\ref{THERMOS_Toolkit_structure}. Calculations for an LTE plasma can be carried out ``ab initio'' in the average atom approximation. Currently, three variants of this model are available: the Thomas-Fermi model (THERMOS\_TF), the modified Hartree-Fock-Slater (THERMOS\_HFS) model, and the quasizone model (THERMOS\_QZM) \cite{Voropinov1970}). In addition, for a low density plasma, the Saha-Boltzmann (THERMOS\_SB) statistics can be used. For the non-LTE cases, the system of level-kinetics equations is solved in the quasi-stationary approach with a fixed ambient radiation field by using the collisional-radiative equilibrium (THERMOS\_CRE) model. The THERMOS\_SB and THERMOS\_CRE models require an external atomic database.

Calculation of the atomic database for a particular substance is based on the isolated-ion model with automatic selection of non-relativistic configurations of ions according to given criteria. The toolkit includes a module that allows to improve the positions of spectral lines and the oscillator strengths by using external data from detailed atomic codes, such as RCG~\cite{Cowan1981} and FAC~\cite{Gu2008}, as well as the known experimental data. In addition to that, a special technique has been developed for averaging the atomic data on a given photon energy grid --- the Radiative Unresolved Spectra Atomic Model or RUSAM~\cite{Novikov2009}, which is aimed at reduction of the calculation time with little to none detriment to the accuracy.

\section{Solid-density aluminum plasma}
\label{subSection_IPD_Al}

The aluminum cases, discussed at the 10$^{th}$ NLTE Workshop, were aimed to explore the collisional-radiative kinetics and the K-shell emission in a solid-density plasma relevant to the XFEL experiments \cite{PhysRevLett.109.065002}. To compare how the density effects are accounted for in different codes, it was proposed to calculate the emissivity of the aluminum plasma at temperatures $T_e=10$, 30, 100, and 300~eV, and the electron number densities $N_e =2\times 10^{23}$ and $5\times 10^{23}$~cm$^{-3}$. To compare THERMOS with other codes, we have selected points with the electron density $N_e = 5 \times 10^{23}$~cm$^{-3}$ and the temperatures $T_e = 30$ and 100~eV.

There is some uncertainty with regard to which of the existing IPD models would be the most appropriate for modeling a dense aluminum plasma \cite{PhysRevLett.109.065002, IGLESIAS2013103}. In the THERMOS toolkit, we have a choice between two IPD models. In the Stewart-Pyatt (SP) approximation \cite{stewart_lowering_1966}, the ionization potential depression $\Delta I$ (in atomic units) for an ion with a net charge $Z$ is given by
$$\Delta I^{SP} (Z) = 3(Z+1)/2r_0,$$
where the dimensionless radius $r_0$ of the electrically neutral atomic cell is defined by the relation $(4\pi/3)(r_0a_0)^3N_i= 1$; here $N_i = N_e/Z$ is the ion number density, $a_0=0.529 \times 10^{-8}$~cm is the Bohr radius. The Ecker-Kr{\"o}ll (EK) approximation \cite{doi:10.1063/1.1724509} uses the expression
$$\Delta I^{EK}(Z)=(Z+1)/r_{EK},$$
where $r_{EK}^3=r_0^3/(1+Z_0)$ is the modified atomic сell radius depending on the mean ion charge $Z_0$.

Initially, the atomic database was calculated for isolated aluminum ions, where only the excited states with the principal quantum numbers $n \leq n_{max} = 6$ were taken into account. Then, as the IPD correction was applied, all the excited states with the binding energies below $\Delta I$ were discarded, which caused $n_{max}$ to decrease. In the present approximation, the radius $r_{EK}$ defines by the ion charge $Z$.

\begin{figure}[!htb]
 \centering
\begingroup
  \makeatletter
  \providecommand\color[2][]{%
    \GenericError{(gnuplot) \space\space\space\@spaces}{%
      Package color not loaded in conjunction with
      terminal option `colourtext'%
    }{See the gnuplot documentation for explanation.%
    }{Either use 'blacktext' in gnuplot or load the package
      color.sty in LaTeX.}%
    \renewcommand\color[2][]{}%
  }%
  \providecommand\includegraphics[2][]{%
    \GenericError{(gnuplot) \space\space\space\@spaces}{%
      Package graphicx or graphics not loaded%
    }{See the gnuplot documentation for explanation.%
    }{The gnuplot epslatex terminal needs graphicx.sty or graphics.sty.}%
    \renewcommand\includegraphics[2][]{}%
  }%
  \providecommand\rotatebox[2]{#2}%
  \@ifundefined{ifGPcolor}{%
    \newif\ifGPcolor
    \GPcolortrue
  }{}%
  \@ifundefined{ifGPblacktext}{%
    \newif\ifGPblacktext
    \GPblacktexttrue
  }{}%
  \let\gplgaddtomacro\g@addto@macro
  \gdef\gplbacktext{}%
  \gdef\gplfronttext{}%
  \makeatother
  \ifGPblacktext
    \def\colorrgb#1{}%
    \def\colorgray#1{}%
  \else
    \ifGPcolor
      \def\colorrgb#1{\color[rgb]{#1}}%
      \def\colorgray#1{\color[gray]{#1}}%
      \expandafter\def\csname LTw\endcsname{\color{white}}%
      \expandafter\def\csname LTb\endcsname{\color{black}}%
      \expandafter\def\csname LTa\endcsname{\color{black}}%
      \expandafter\def\csname LT0\endcsname{\color[rgb]{1,0,0}}%
      \expandafter\def\csname LT1\endcsname{\color[rgb]{0,1,0}}%
      \expandafter\def\csname LT2\endcsname{\color[rgb]{0,0,1}}%
      \expandafter\def\csname LT3\endcsname{\color[rgb]{1,0,1}}%
      \expandafter\def\csname LT4\endcsname{\color[rgb]{0,1,1}}%
      \expandafter\def\csname LT5\endcsname{\color[rgb]{1,1,0}}%
      \expandafter\def\csname LT6\endcsname{\color[rgb]{0,0,0}}%
      \expandafter\def\csname LT7\endcsname{\color[rgb]{1,0.3,0}}%
      \expandafter\def\csname LT8\endcsname{\color[rgb]{0.5,0.5,0.5}}%
    \else
      \def\colorrgb#1{\color{black}}%
      \def\colorgray#1{\color[gray]{#1}}%
      \expandafter\def\csname LTw\endcsname{\color{white}}%
      \expandafter\def\csname LTb\endcsname{\color{black}}%
      \expandafter\def\csname LTa\endcsname{\color{black}}%
      \expandafter\def\csname LT0\endcsname{\color{black}}%
      \expandafter\def\csname LT1\endcsname{\color{black}}%
      \expandafter\def\csname LT2\endcsname{\color{black}}%
      \expandafter\def\csname LT3\endcsname{\color{black}}%
      \expandafter\def\csname LT4\endcsname{\color{black}}%
      \expandafter\def\csname LT5\endcsname{\color{black}}%
      \expandafter\def\csname LT6\endcsname{\color{black}}%
      \expandafter\def\csname LT7\endcsname{\color{black}}%
      \expandafter\def\csname LT8\endcsname{\color{black}}%
    \fi
  \fi
  \setlength{\unitlength}{0.0500bp}%
  \begin{picture}(5102.00,3400.00)%
    \gplgaddtomacro\gplbacktext{%
      \csname LTb\endcsname%
      \put(902,704){\makebox(0,0)[r]{\strut{} 1}}%
      \csname LTb\endcsname%
      \put(902,1312){\makebox(0,0)[r]{\strut{} 2}}%
      \csname LTb\endcsname%
      \put(902,1920){\makebox(0,0)[r]{\strut{} 3}}%
      \csname LTb\endcsname%
      \put(902,2527){\makebox(0,0)[r]{\strut{} 4}}%
      \csname LTb\endcsname%
      \put(902,3135){\makebox(0,0)[r]{\strut{} 5}}%
      \csname LTb\endcsname%
      \put(1034,484){\makebox(0,0){\strut{} 0}}%
      \csname LTb\endcsname%
      \put(1646,484){\makebox(0,0){\strut{} 2}}%
      \csname LTb\endcsname%
      \put(2258,484){\makebox(0,0){\strut{} 4}}%
      \csname LTb\endcsname%
      \put(2870,484){\makebox(0,0){\strut{} 6}}%
      \csname LTb\endcsname%
      \put(3481,484){\makebox(0,0){\strut{} 8}}%
      \csname LTb\endcsname%
      \put(4093,484){\makebox(0,0){\strut{} 10}}%
      \csname LTb\endcsname%
      \put(4705,484){\makebox(0,0){\strut{} 12}}%
      \put(176,1919){\rotatebox{-270}{\makebox(0,0){\strut{}Principal quantum number }}}%
      \put(396,1919){\rotatebox{-270}{\makebox(0,0){\strut{} of the highest states}}}%
      \put(2869,154){\makebox(0,0){\strut{}Ion charge}}%
    }%
    \gplgaddtomacro\gplfronttext{%
      \csname LTb\endcsname%
      \put(2021,2962){\makebox(0,0)[l]{\strut{}THERMOS\_CRE SP}}%
      \csname LTb\endcsname%
      \put(2021,2742){\makebox(0,0)[l]{\strut{}THERMOS\_CRE EK}}%
    }%
    \gplbacktext
    \put(0,0){\includegraphics{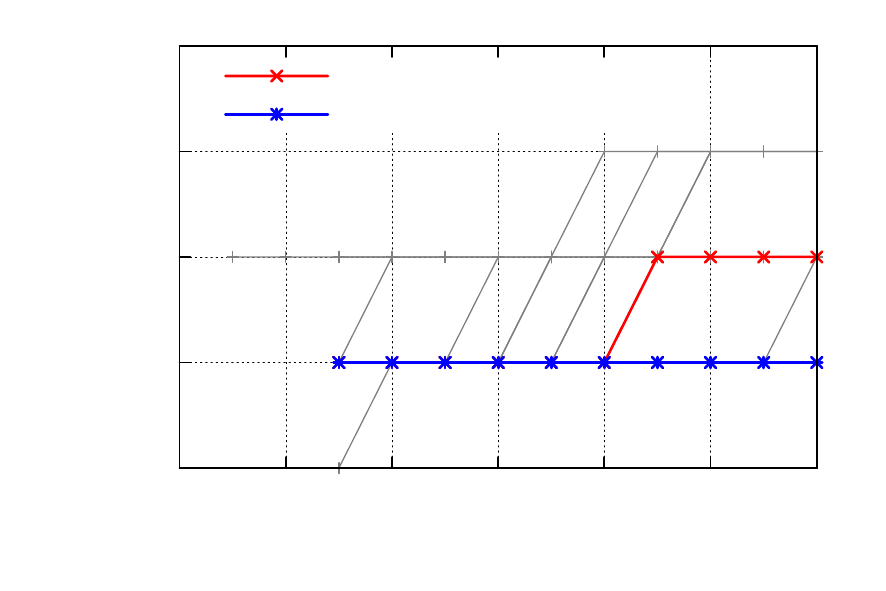}}%
    \gplfronttext
  \end{picture}%
\endgroup
\caption{Maximum principal quantum number $n_{max}$ versus the ion charge $Z$ in the aluminum plasma with the electron density $5\times 10^{23}$~cm$^{-3}$ and temperature 100~eV.}
\label{Alnmax_100}
\end{figure}

The calculated dependence of $n_{max}$ on the ion charge $Z$ for the electron density $N_e = 5 \times 10^{23}$~cm$^{-3}$ and temperature $T_e=100$~eV is shown in Fig.~\ref{Alnmax_100}. This figure shows also the results by other participants of the 10$^{th}$ NLTE Workshop (gray color). One observes a broad scatter of the results. The THERMOS toolkit yields $n_{max}=2$ for the ions with charges $3\leq Z <9$ within the Stewart-Pyatt approximation, and $n_{max}= 2$ for all the ions $Z \geq 3$ when the Ecker-Kr{\"o}ll formula is used. Ions with $Z < 3$ are not present under the considered conditions. Both approximations produce very close results, and we cannot give preference to any of them.

\begin{figure}[!htb]
 \centering
\begingroup
  \makeatletter
  \providecommand\color[2][]{%
    \GenericError{(gnuplot) \space\space\space\@spaces}{%
      Package color not loaded in conjunction with
      terminal option `colourtext'%
    }{See the gnuplot documentation for explanation.%
    }{Either use 'blacktext' in gnuplot or load the package
      color.sty in LaTeX.}%
    \renewcommand\color[2][]{}%
  }%
  \providecommand\includegraphics[2][]{%
    \GenericError{(gnuplot) \space\space\space\@spaces}{%
      Package graphicx or graphics not loaded%
    }{See the gnuplot documentation for explanation.%
    }{The gnuplot epslatex terminal needs graphicx.sty or graphics.sty.}%
    \renewcommand\includegraphics[2][]{}%
  }%
  \providecommand\rotatebox[2]{#2}%
  \@ifundefined{ifGPcolor}{%
    \newif\ifGPcolor
    \GPcolortrue
  }{}%
  \@ifundefined{ifGPblacktext}{%
    \newif\ifGPblacktext
    \GPblacktexttrue
  }{}%
  \let\gplgaddtomacro\g@addto@macro
  \gdef\gplbacktext{}%
  \gdef\gplfronttext{}%
  \makeatother
  \ifGPblacktext
    \def\colorrgb#1{}%
    \def\colorgray#1{}%
  \else
    \ifGPcolor
      \def\colorrgb#1{\color[rgb]{#1}}%
      \def\colorgray#1{\color[gray]{#1}}%
      \expandafter\def\csname LTw\endcsname{\color{white}}%
      \expandafter\def\csname LTb\endcsname{\color{black}}%
      \expandafter\def\csname LTa\endcsname{\color{black}}%
      \expandafter\def\csname LT0\endcsname{\color[rgb]{1,0,0}}%
      \expandafter\def\csname LT1\endcsname{\color[rgb]{0,1,0}}%
      \expandafter\def\csname LT2\endcsname{\color[rgb]{0,0,1}}%
      \expandafter\def\csname LT3\endcsname{\color[rgb]{1,0,1}}%
      \expandafter\def\csname LT4\endcsname{\color[rgb]{0,1,1}}%
      \expandafter\def\csname LT5\endcsname{\color[rgb]{1,1,0}}%
      \expandafter\def\csname LT6\endcsname{\color[rgb]{0,0,0}}%
      \expandafter\def\csname LT7\endcsname{\color[rgb]{1,0.3,0}}%
      \expandafter\def\csname LT8\endcsname{\color[rgb]{0.5,0.5,0.5}}%
    \else
      \def\colorrgb#1{\color{black}}%
      \def\colorgray#1{\color[gray]{#1}}%
      \expandafter\def\csname LTw\endcsname{\color{white}}%
      \expandafter\def\csname LTb\endcsname{\color{black}}%
      \expandafter\def\csname LTa\endcsname{\color{black}}%
      \expandafter\def\csname LT0\endcsname{\color{black}}%
      \expandafter\def\csname LT1\endcsname{\color{black}}%
      \expandafter\def\csname LT2\endcsname{\color{black}}%
      \expandafter\def\csname LT3\endcsname{\color{black}}%
      \expandafter\def\csname LT4\endcsname{\color{black}}%
      \expandafter\def\csname LT5\endcsname{\color{black}}%
      \expandafter\def\csname LT6\endcsname{\color{black}}%
      \expandafter\def\csname LT7\endcsname{\color{black}}%
      \expandafter\def\csname LT8\endcsname{\color{black}}%
    \fi
  \fi
  \setlength{\unitlength}{0.0500bp}%
  \begin{picture}(5102.00,3400.00)%
    \gplgaddtomacro\gplbacktext{%
      \csname LTb\endcsname%
      \put(726,704){\makebox(0,0)[r]{\strut{}10$^{9}$}}%
      \csname LTb\endcsname%
      \put(726,1312){\makebox(0,0)[r]{\strut{}10$^{10}$}}%
      \csname LTb\endcsname%
      \put(726,1920){\makebox(0,0)[r]{\strut{}10$^{11}$}}%
      \csname LTb\endcsname%
      \put(726,2527){\makebox(0,0)[r]{\strut{}10$^{12}$}}%
      \csname LTb\endcsname%
      \put(726,3135){\makebox(0,0)[r]{\strut{}10$^{13}$}}%
      \csname LTb\endcsname%
      \put(858,484){\makebox(0,0){\strut{} 1400}}%
      \csname LTb\endcsname%
      \put(1499,484){\makebox(0,0){\strut{} 1500}}%
      \csname LTb\endcsname%
      \put(2140,484){\makebox(0,0){\strut{} 1600}}%
      \csname LTb\endcsname%
      \put(2782,484){\makebox(0,0){\strut{} 1700}}%
      \csname LTb\endcsname%
      \put(3423,484){\makebox(0,0){\strut{} 1800}}%
      \csname LTb\endcsname%
      \put(4064,484){\makebox(0,0){\strut{} 1900}}%
      \csname LTb\endcsname%
      \put(4705,484){\makebox(0,0){\strut{} 2000}}%
      \put(220,1919){\rotatebox{-270}{\makebox(0,0){\strut{}Emissivity, J/s/cm$^3$/eV}}}%
      \put(2781,154){\makebox(0,0){\strut{}Photon energy $\omega$, eV}}%
    }%
    \gplgaddtomacro\gplfronttext{%
      \csname LTb\endcsname%
      \put(3718,2962){\makebox(0,0)[r]{\strut{}THERMOS\_CRE SP}}%
    }%
    \gplbacktext
    \put(0,0){\includegraphics{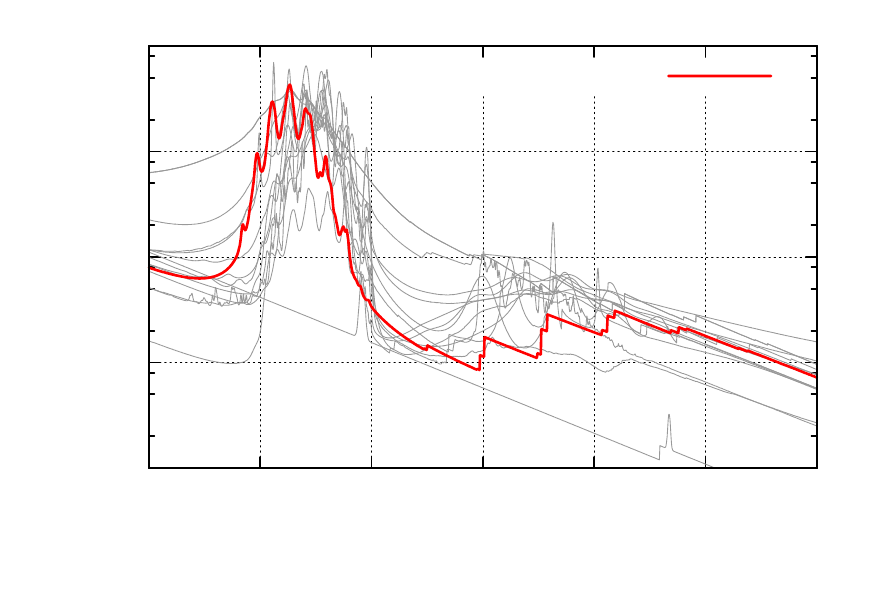}}%
    \gplfronttext
  \end{picture}%
\endgroup
\caption{Emissivity of the aluminum plasma with the electron density $5\times 10^{23}$~cm$^{-3}$ and temperature 100~eV.}
\label{Al_NLTEb}
\end{figure}

\begin{figure}[!htb]
 \centering
\begingroup
  \makeatletter
  \providecommand\color[2][]{%
    \GenericError{(gnuplot) \space\space\space\@spaces}{%
      Package color not loaded in conjunction with
      terminal option `colourtext'%
    }{See the gnuplot documentation for explanation.%
    }{Either use 'blacktext' in gnuplot or load the package
      color.sty in LaTeX.}%
    \renewcommand\color[2][]{}%
  }%
  \providecommand\includegraphics[2][]{%
    \GenericError{(gnuplot) \space\space\space\@spaces}{%
      Package graphicx or graphics not loaded%
    }{See the gnuplot documentation for explanation.%
    }{The gnuplot epslatex terminal needs graphicx.sty or graphics.sty.}%
    \renewcommand\includegraphics[2][]{}%
  }%
  \providecommand\rotatebox[2]{#2}%
  \@ifundefined{ifGPcolor}{%
    \newif\ifGPcolor
    \GPcolortrue
  }{}%
  \@ifundefined{ifGPblacktext}{%
    \newif\ifGPblacktext
    \GPblacktexttrue
  }{}%
  \let\gplgaddtomacro\g@addto@macro
  \gdef\gplbacktext{}%
  \gdef\gplfronttext{}%
  \makeatother
  \ifGPblacktext
    \def\colorrgb#1{}%
    \def\colorgray#1{}%
  \else
    \ifGPcolor
      \def\colorrgb#1{\color[rgb]{#1}}%
      \def\colorgray#1{\color[gray]{#1}}%
      \expandafter\def\csname LTw\endcsname{\color{white}}%
      \expandafter\def\csname LTb\endcsname{\color{black}}%
      \expandafter\def\csname LTa\endcsname{\color{black}}%
      \expandafter\def\csname LT0\endcsname{\color[rgb]{1,0,0}}%
      \expandafter\def\csname LT1\endcsname{\color[rgb]{0,1,0}}%
      \expandafter\def\csname LT2\endcsname{\color[rgb]{0,0,1}}%
      \expandafter\def\csname LT3\endcsname{\color[rgb]{1,0,1}}%
      \expandafter\def\csname LT4\endcsname{\color[rgb]{0,1,1}}%
      \expandafter\def\csname LT5\endcsname{\color[rgb]{1,1,0}}%
      \expandafter\def\csname LT6\endcsname{\color[rgb]{0,0,0}}%
      \expandafter\def\csname LT7\endcsname{\color[rgb]{1,0.3,0}}%
      \expandafter\def\csname LT8\endcsname{\color[rgb]{0.5,0.5,0.5}}%
    \else
      \def\colorrgb#1{\color{black}}%
      \def\colorgray#1{\color[gray]{#1}}%
      \expandafter\def\csname LTw\endcsname{\color{white}}%
      \expandafter\def\csname LTb\endcsname{\color{black}}%
      \expandafter\def\csname LTa\endcsname{\color{black}}%
      \expandafter\def\csname LT0\endcsname{\color{black}}%
      \expandafter\def\csname LT1\endcsname{\color{black}}%
      \expandafter\def\csname LT2\endcsname{\color{black}}%
      \expandafter\def\csname LT3\endcsname{\color{black}}%
      \expandafter\def\csname LT4\endcsname{\color{black}}%
      \expandafter\def\csname LT5\endcsname{\color{black}}%
      \expandafter\def\csname LT6\endcsname{\color{black}}%
      \expandafter\def\csname LT7\endcsname{\color{black}}%
      \expandafter\def\csname LT8\endcsname{\color{black}}%
    \fi
  \fi
  \setlength{\unitlength}{0.0500bp}%
  \begin{picture}(5102.00,3400.00)%
    \gplgaddtomacro\gplbacktext{%
      \csname LTb\endcsname%
      \put(726,704){\makebox(0,0)[r]{\strut{} 0}}%
      \csname LTb\endcsname%
      \put(726,1109){\makebox(0,0)[r]{\strut{} 0.1}}%
      \csname LTb\endcsname%
      \put(726,1514){\makebox(0,0)[r]{\strut{} 0.2}}%
      \csname LTb\endcsname%
      \put(726,1920){\makebox(0,0)[r]{\strut{} 0.3}}%
      \csname LTb\endcsname%
      \put(726,2325){\makebox(0,0)[r]{\strut{} 0.4}}%
      \csname LTb\endcsname%
      \put(726,2730){\makebox(0,0)[r]{\strut{} 0.5}}%
      \csname LTb\endcsname%
      \put(726,3135){\makebox(0,0)[r]{\strut{} 0.6}}%
      \csname LTb\endcsname%
      \put(858,484){\makebox(0,0){\strut{} 3}}%
      \csname LTb\endcsname%
      \put(1339,484){\makebox(0,0){\strut{} 4}}%
      \csname LTb\endcsname%
      \put(1820,484){\makebox(0,0){\strut{} 5}}%
      \csname LTb\endcsname%
      \put(2301,484){\makebox(0,0){\strut{} 6}}%
      \csname LTb\endcsname%
      \put(2782,484){\makebox(0,0){\strut{} 7}}%
      \csname LTb\endcsname%
      \put(3262,484){\makebox(0,0){\strut{} 8}}%
      \csname LTb\endcsname%
      \put(3743,484){\makebox(0,0){\strut{} 9}}%
      \csname LTb\endcsname%
      \put(4224,484){\makebox(0,0){\strut{} 10}}%
      \csname LTb\endcsname%
      \put(4705,484){\makebox(0,0){\strut{} 11}}%
      \put(220,1919){\rotatebox{-270}{\makebox(0,0){\strut{}Ion population}}}%
      \put(2781,154){\makebox(0,0){\strut{}Ion charge}}%
    }%
    \gplgaddtomacro\gplfronttext{%
      \csname LTb\endcsname%
      \put(3718,2962){\makebox(0,0)[r]{\strut{}THERMOS\_CRE}}%
    }%
    \gplbacktext
    \put(0,0){\includegraphics{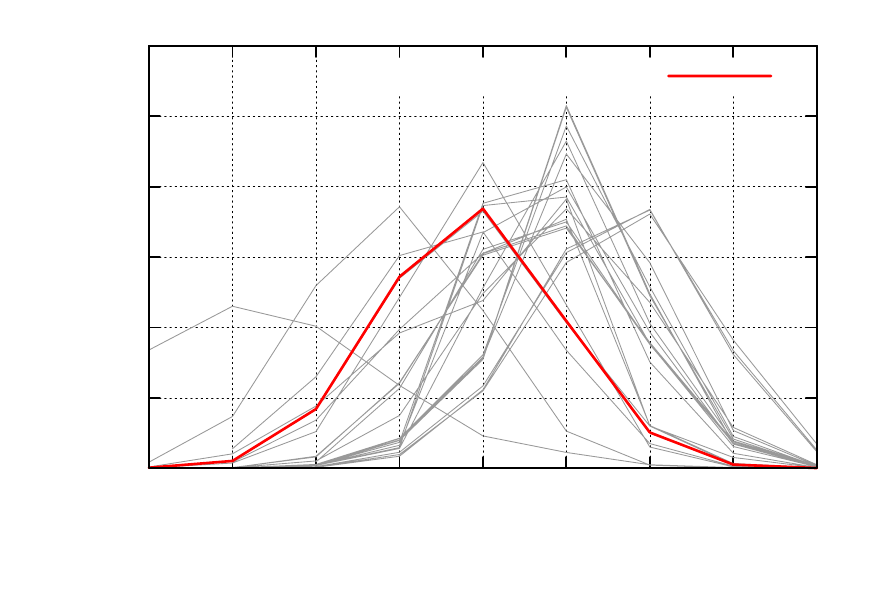}}%
    \gplfronttext
  \end{picture}%
\endgroup
\caption{Ion populations versus the ion charge in the aluminum plasma with the electron density $5\times 10^{23}$~cm$^{-3}$ and temperature~100~eV. Red line shows the THERMOS\_CRE results with the IPD by Stewart-Pyatt at $Z_0=6.85$.}
 \label{Alpop_b}
\end{figure}

Figure ~\ref{Al_NLTEb} shows the emissivity of the aluminum plasma, calculated with the THERMOS toolkit, in comparison with the other results from the 10$^{th}$ NLTE Workshop. Generally,  a good agreement with the results by other scientific groups is observed, although the THERMOS spectrum is shifted towards lower photon energies due to the lower mean ion charge (see~Fig.~\ref{Alpop_b}).

\begin{figure}[!htb]
 \centering
\begingroup
  \makeatletter
  \providecommand\color[2][]{%
    \GenericError{(gnuplot) \space\space\space\@spaces}{%
      Package color not loaded in conjunction with
      terminal option `colourtext'%
    }{See the gnuplot documentation for explanation.%
    }{Either use 'blacktext' in gnuplot or load the package
      color.sty in LaTeX.}%
    \renewcommand\color[2][]{}%
  }%
  \providecommand\includegraphics[2][]{%
    \GenericError{(gnuplot) \space\space\space\@spaces}{%
      Package graphicx or graphics not loaded%
    }{See the gnuplot documentation for explanation.%
    }{The gnuplot epslatex terminal needs graphicx.sty or graphics.sty.}%
    \renewcommand\includegraphics[2][]{}%
  }%
  \providecommand\rotatebox[2]{#2}%
  \@ifundefined{ifGPcolor}{%
    \newif\ifGPcolor
    \GPcolortrue
  }{}%
  \@ifundefined{ifGPblacktext}{%
    \newif\ifGPblacktext
    \GPblacktexttrue
  }{}%
  \let\gplgaddtomacro\g@addto@macro
  \gdef\gplbacktext{}%
  \gdef\gplfronttext{}%
  \makeatother
  \ifGPblacktext
    \def\colorrgb#1{}%
    \def\colorgray#1{}%
  \else
    \ifGPcolor
      \def\colorrgb#1{\color[rgb]{#1}}%
      \def\colorgray#1{\color[gray]{#1}}%
      \expandafter\def\csname LTw\endcsname{\color{white}}%
      \expandafter\def\csname LTb\endcsname{\color{black}}%
      \expandafter\def\csname LTa\endcsname{\color{black}}%
      \expandafter\def\csname LT0\endcsname{\color[rgb]{1,0,0}}%
      \expandafter\def\csname LT1\endcsname{\color[rgb]{0,1,0}}%
      \expandafter\def\csname LT2\endcsname{\color[rgb]{0,0,1}}%
      \expandafter\def\csname LT3\endcsname{\color[rgb]{1,0,1}}%
      \expandafter\def\csname LT4\endcsname{\color[rgb]{0,1,1}}%
      \expandafter\def\csname LT5\endcsname{\color[rgb]{1,1,0}}%
      \expandafter\def\csname LT6\endcsname{\color[rgb]{0,0,0}}%
      \expandafter\def\csname LT7\endcsname{\color[rgb]{1,0.3,0}}%
      \expandafter\def\csname LT8\endcsname{\color[rgb]{0.5,0.5,0.5}}%
    \else
      \def\colorrgb#1{\color{black}}%
      \def\colorgray#1{\color[gray]{#1}}%
      \expandafter\def\csname LTw\endcsname{\color{white}}%
      \expandafter\def\csname LTb\endcsname{\color{black}}%
      \expandafter\def\csname LTa\endcsname{\color{black}}%
      \expandafter\def\csname LT0\endcsname{\color{black}}%
      \expandafter\def\csname LT1\endcsname{\color{black}}%
      \expandafter\def\csname LT2\endcsname{\color{black}}%
      \expandafter\def\csname LT3\endcsname{\color{black}}%
      \expandafter\def\csname LT4\endcsname{\color{black}}%
      \expandafter\def\csname LT5\endcsname{\color{black}}%
      \expandafter\def\csname LT6\endcsname{\color{black}}%
      \expandafter\def\csname LT7\endcsname{\color{black}}%
      \expandafter\def\csname LT8\endcsname{\color{black}}%
    \fi
  \fi
  \setlength{\unitlength}{0.0500bp}%
  \begin{picture}(5102.00,3400.00)%
    \gplgaddtomacro\gplbacktext{%
      \csname LTb\endcsname%
      \put(726,704){\makebox(0,0)[r]{\strut{} 0}}%
      \csname LTb\endcsname%
      \put(726,1190){\makebox(0,0)[r]{\strut{} 0.2}}%
      \csname LTb\endcsname%
      \put(726,1676){\makebox(0,0)[r]{\strut{} 0.4}}%
      \csname LTb\endcsname%
      \put(726,2163){\makebox(0,0)[r]{\strut{} 0.6}}%
      \csname LTb\endcsname%
      \put(726,2649){\makebox(0,0)[r]{\strut{} 0.8}}%
      \csname LTb\endcsname%
      \put(726,3135){\makebox(0,0)[r]{\strut{} 1}}%
      \csname LTb\endcsname%
      \put(858,484){\makebox(0,0){\strut{} 3}}%
      \csname LTb\endcsname%
      \put(1627,484){\makebox(0,0){\strut{} 4}}%
      \csname LTb\endcsname%
      \put(2397,484){\makebox(0,0){\strut{} 5}}%
      \csname LTb\endcsname%
      \put(3166,484){\makebox(0,0){\strut{} 6}}%
      \csname LTb\endcsname%
      \put(3936,484){\makebox(0,0){\strut{} 7}}%
      \csname LTb\endcsname%
      \put(4705,484){\makebox(0,0){\strut{} 8}}%
      \put(220,1919){\rotatebox{-270}{\makebox(0,0){\strut{}Ion population}}}%
      \put(2781,154){\makebox(0,0){\strut{}Ion charge}}%
    }%
    \gplgaddtomacro\gplfronttext{%
      \csname LTb\endcsname%
      \put(3718,2962){\makebox(0,0)[r]{\strut{}THERMOS\_HFS}}%
      \csname LTb\endcsname%
      \put(3718,2742){\makebox(0,0)[r]{\strut{}THERMOS\_CRE}}%
    }%
    \gplbacktext
    \put(0,0){\includegraphics{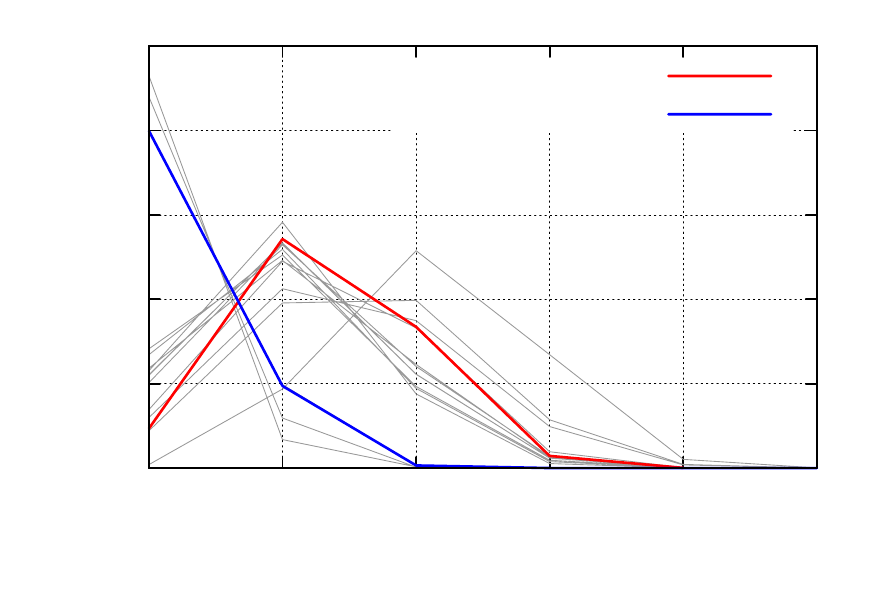}}%
    \gplfronttext
  \end{picture}%
\endgroup
\caption{Ion populations versus the ion charge in the aluminum plasma with the electron density $5\times 10^{23}$~cm$^{-3}$ and temperature~30~eV.}
\label{Al30Pk_LTE}
\end{figure}

\begin{figure}[!htb]
 \centering
\begingroup
  \makeatletter
  \providecommand\color[2][]{%
    \GenericError{(gnuplot) \space\space\space\@spaces}{%
      Package color not loaded in conjunction with
      terminal option `colourtext'%
    }{See the gnuplot documentation for explanation.%
    }{Either use 'blacktext' in gnuplot or load the package
      color.sty in LaTeX.}%
    \renewcommand\color[2][]{}%
  }%
  \providecommand\includegraphics[2][]{%
    \GenericError{(gnuplot) \space\space\space\@spaces}{%
      Package graphicx or graphics not loaded%
    }{See the gnuplot documentation for explanation.%
    }{The gnuplot epslatex terminal needs graphicx.sty or graphics.sty.}%
    \renewcommand\includegraphics[2][]{}%
  }%
  \providecommand\rotatebox[2]{#2}%
  \@ifundefined{ifGPcolor}{%
    \newif\ifGPcolor
    \GPcolortrue
  }{}%
  \@ifundefined{ifGPblacktext}{%
    \newif\ifGPblacktext
    \GPblacktexttrue
  }{}%
  \let\gplgaddtomacro\g@addto@macro
  \gdef\gplbacktext{}%
  \gdef\gplfronttext{}%
  \makeatother
  \ifGPblacktext
    \def\colorrgb#1{}%
    \def\colorgray#1{}%
  \else
    \ifGPcolor
      \def\colorrgb#1{\color[rgb]{#1}}%
      \def\colorgray#1{\color[gray]{#1}}%
      \expandafter\def\csname LTw\endcsname{\color{white}}%
      \expandafter\def\csname LTb\endcsname{\color{black}}%
      \expandafter\def\csname LTa\endcsname{\color{black}}%
      \expandafter\def\csname LT0\endcsname{\color[rgb]{1,0,0}}%
      \expandafter\def\csname LT1\endcsname{\color[rgb]{0,1,0}}%
      \expandafter\def\csname LT2\endcsname{\color[rgb]{0,0,1}}%
      \expandafter\def\csname LT3\endcsname{\color[rgb]{1,0,1}}%
      \expandafter\def\csname LT4\endcsname{\color[rgb]{0,1,1}}%
      \expandafter\def\csname LT5\endcsname{\color[rgb]{1,1,0}}%
      \expandafter\def\csname LT6\endcsname{\color[rgb]{0,0,0}}%
      \expandafter\def\csname LT7\endcsname{\color[rgb]{1,0.3,0}}%
      \expandafter\def\csname LT8\endcsname{\color[rgb]{0.5,0.5,0.5}}%
    \else
      \def\colorrgb#1{\color{black}}%
      \def\colorgray#1{\color[gray]{#1}}%
      \expandafter\def\csname LTw\endcsname{\color{white}}%
      \expandafter\def\csname LTb\endcsname{\color{black}}%
      \expandafter\def\csname LTa\endcsname{\color{black}}%
      \expandafter\def\csname LT0\endcsname{\color{black}}%
      \expandafter\def\csname LT1\endcsname{\color{black}}%
      \expandafter\def\csname LT2\endcsname{\color{black}}%
      \expandafter\def\csname LT3\endcsname{\color{black}}%
      \expandafter\def\csname LT4\endcsname{\color{black}}%
      \expandafter\def\csname LT5\endcsname{\color{black}}%
      \expandafter\def\csname LT6\endcsname{\color{black}}%
      \expandafter\def\csname LT7\endcsname{\color{black}}%
      \expandafter\def\csname LT8\endcsname{\color{black}}%
    \fi
  \fi
  \setlength{\unitlength}{0.0500bp}%
  \begin{picture}(5102.00,3400.00)%
    \gplgaddtomacro\gplbacktext{%
      \csname LTb\endcsname%
      \put(814,704){\makebox(0,0)[r]{\strut{}10$^{-9}$}}%
      \csname LTb\endcsname%
      \put(814,1312){\makebox(0,0)[r]{\strut{}10$^{-7}$}}%
      \csname LTb\endcsname%
      \put(814,1920){\makebox(0,0)[r]{\strut{}10$^{-5}$}}%
      \csname LTb\endcsname%
      \put(814,2527){\makebox(0,0)[r]{\strut{}10$^{-3}$}}%
      \csname LTb\endcsname%
      \put(814,3135){\makebox(0,0)[r]{\strut{}10$^{-1}$}}%
      \csname LTb\endcsname%
      \put(946,484){\makebox(0,0){\strut{} 1400}}%
      \csname LTb\endcsname%
      \put(1886,484){\makebox(0,0){\strut{} 1500}}%
      \csname LTb\endcsname%
      \put(2826,484){\makebox(0,0){\strut{} 1600}}%
      \csname LTb\endcsname%
      \put(3765,484){\makebox(0,0){\strut{} 1700}}%
      \csname LTb\endcsname%
      \put(4705,484){\makebox(0,0){\strut{} 1800}}%
      \put(176,1919){\rotatebox{-270}{\makebox(0,0){\strut{}Emissivity, J/s/cm$^3$/eV}}}%
      \put(2825,154){\makebox(0,0){\strut{}Photon energy $\omega$, eV}}%
    }%
    \gplgaddtomacro\gplfronttext{%
      \csname LTb\endcsname%
      \put(3718,2962){\makebox(0,0)[r]{\strut{}THERMOS\_HFS}}%
      \csname LTb\endcsname%
      \put(3718,2742){\makebox(0,0)[r]{\strut{}THERMOS\_CRE}}%
    }%
    \gplbacktext
    \put(0,0){\includegraphics{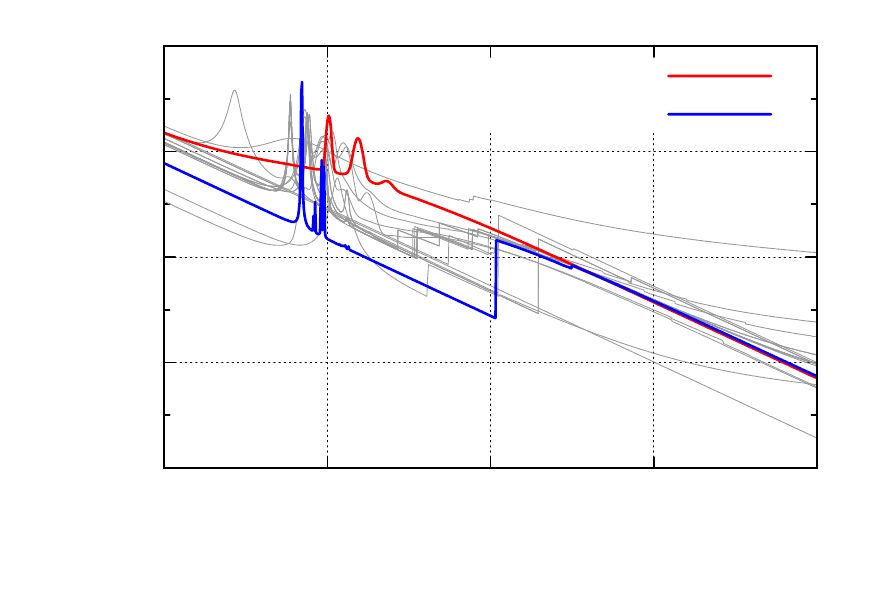}}%
    \gplfronttext
  \end{picture}%
\endgroup
\caption{Emissivity of the aluminum plasma with the electron density $5\times 10^{23}$~cm$^{-3}$ and temperature~30~eV.}
\label{Al30sp_LTE}
\end{figure}

To clarify the cause of the above discrepancy, we examined the case with the electron density $N_e=5\times 10^{23} $~cm$^{-3}$ and the temperature $T_e= 30$~eV. Under these conditions the plasma can be treated as LTE, and the THERMOS\_HFS code is applicable for performing the reference calculation. As is seen in Fig.~\ref{Al30Pk_LTE}, whereas the THERMOS\_CRE produced a mean charge of $Z_0= 3.2$, the THERMOS\_HFS yielded $Z_0= 4.3$ --- which is closer to the 10$^{th}$ NLTE results. The emissivity of the aluminum plasma, calculated with the THERMOS\_CRE and THERMOS\_HFS and compared with other results from the 10$^{th}$ NLTE Workshop, is displayed in Fig.~\ref{Al30sp_LTE}. This figure indicates that the THERMOS\_HFS code does not describe the line positions correctly, although the bound-free cross-sections seem to be in order; also, the broadening of the photoabsorption edges due to the density effects is clearly seen in contrast to the corresponding THERMOS\_CRE results. Here we conclude that in solid-density plasmas the simplest approximations \cite{NovikovRalchenkoBook} for the ionization and recombination rates, used in the present THERMOS\_CRE version, are not applicable and need to be reconsidered.

\section{ORION experiment with the chlorine plasma}
\label{subSection_ORION_Cl}

The simulation of the emissivity of the chlorine plasma was related to the recent high-resolution measurements on the OHREX spectrometer fielded at the Orion facility, as is shown in Fig.~5(b) of Ref.~\cite{doi:10.1063/1.4965233}. According to the experimental data, it was not possible to obtain accurate values for the temperature and density of the plasma, though the ranges of their variation were evaluated as $T_e \approx 400$--600~eV and $N_e \approx 10^{21}$--$10^{23}$~cm$^{-3}$. At the Workshop it was proposed to determine the best fit to the experiment. To reach this goal, it was necessary to calculate the emissivity and to investigate the widths and intensities of the K-lines in a dense chlorine plasma within the given ranges of temperature and density.

\begin{figure}[!htb]
 \centering
\begingroup
  \makeatletter
  \providecommand\color[2][]{%
    \GenericError{(gnuplot) \space\space\space\@spaces}{%
      Package color not loaded in conjunction with
      terminal option `colourtext'%
    }{See the gnuplot documentation for explanation.%
    }{Either use 'blacktext' in gnuplot or load the package
      color.sty in LaTeX.}%
    \renewcommand\color[2][]{}%
  }%
  \providecommand\includegraphics[2][]{%
    \GenericError{(gnuplot) \space\space\space\@spaces}{%
      Package graphicx or graphics not loaded%
    }{See the gnuplot documentation for explanation.%
    }{The gnuplot epslatex terminal needs graphicx.sty or graphics.sty.}%
    \renewcommand\includegraphics[2][]{}%
  }%
  \providecommand\rotatebox[2]{#2}%
  \@ifundefined{ifGPcolor}{%
    \newif\ifGPcolor
    \GPcolortrue
  }{}%
  \@ifundefined{ifGPblacktext}{%
    \newif\ifGPblacktext
    \GPblacktexttrue
  }{}%
  \let\gplgaddtomacro\g@addto@macro
  \gdef\gplbacktext{}%
  \gdef\gplfronttext{}%
  \makeatother
  \ifGPblacktext
    \def\colorrgb#1{}%
    \def\colorgray#1{}%
  \else
    \ifGPcolor
      \def\colorrgb#1{\color[rgb]{#1}}%
      \def\colorgray#1{\color[gray]{#1}}%
      \expandafter\def\csname LTw\endcsname{\color{white}}%
      \expandafter\def\csname LTb\endcsname{\color{black}}%
      \expandafter\def\csname LTa\endcsname{\color{black}}%
      \expandafter\def\csname LT0\endcsname{\color[rgb]{1,0,0}}%
      \expandafter\def\csname LT1\endcsname{\color[rgb]{0,1,0}}%
      \expandafter\def\csname LT2\endcsname{\color[rgb]{0,0,1}}%
      \expandafter\def\csname LT3\endcsname{\color[rgb]{1,0,1}}%
      \expandafter\def\csname LT4\endcsname{\color[rgb]{0,1,1}}%
      \expandafter\def\csname LT5\endcsname{\color[rgb]{1,1,0}}%
      \expandafter\def\csname LT6\endcsname{\color[rgb]{0,0,0}}%
      \expandafter\def\csname LT7\endcsname{\color[rgb]{1,0.3,0}}%
      \expandafter\def\csname LT8\endcsname{\color[rgb]{0.5,0.5,0.5}}%
    \else
      \def\colorrgb#1{\color{black}}%
      \def\colorgray#1{\color[gray]{#1}}%
      \expandafter\def\csname LTw\endcsname{\color{white}}%
      \expandafter\def\csname LTb\endcsname{\color{black}}%
      \expandafter\def\csname LTa\endcsname{\color{black}}%
      \expandafter\def\csname LT0\endcsname{\color{black}}%
      \expandafter\def\csname LT1\endcsname{\color{black}}%
      \expandafter\def\csname LT2\endcsname{\color{black}}%
      \expandafter\def\csname LT3\endcsname{\color{black}}%
      \expandafter\def\csname LT4\endcsname{\color{black}}%
      \expandafter\def\csname LT5\endcsname{\color{black}}%
      \expandafter\def\csname LT6\endcsname{\color{black}}%
      \expandafter\def\csname LT7\endcsname{\color{black}}%
      \expandafter\def\csname LT8\endcsname{\color{black}}%
    \fi
  \fi
    \setlength{\unitlength}{0.0500bp}%
    \ifx\gptboxheight\undefined%
      \newlength{\gptboxheight}%
      \newlength{\gptboxwidth}%
      \newsavebox{\gptboxtext}%
    \fi%
    \setlength{\fboxrule}{0.5pt}%
    \setlength{\fboxsep}{1pt}%
\begin{picture}(5102.00,3400.00)%
    \gplgaddtomacro\gplbacktext{%
      \csname LTb\endcsname
      \put(814,704){\makebox(0,0)[r]{\strut{}$0$}}%
      \csname LTb\endcsname
      \put(814,1199){\makebox(0,0)[r]{\strut{}$0.2$}}%
      \csname LTb\endcsname
      \put(814,1694){\makebox(0,0)[r]{\strut{}$0.4$}}%
      \csname LTb\endcsname
      \put(814,2189){\makebox(0,0)[r]{\strut{}$0.6$}}%
      \csname LTb\endcsname
      \put(814,2684){\makebox(0,0)[r]{\strut{}$0.8$}}%
      \csname LTb\endcsname
      \put(814,3179){\makebox(0,0)[r]{\strut{}$1$}}%
      \csname LTb\endcsname
      \put(946,484){\makebox(0,0){\strut{}$3150$}}%
      \csname LTb\endcsname
      \put(1886,484){\makebox(0,0){\strut{}$3200$}}%
      \csname LTb\endcsname
      \put(2826,484){\makebox(0,0){\strut{}$3250$}}%
      \csname LTb\endcsname
      \put(3765,484){\makebox(0,0){\strut{}$3300$}}%
      \csname LTb\endcsname
      \put(4705,484){\makebox(0,0){\strut{}$3350$}}%
    }%
    \gplgaddtomacro\gplfronttext{%
      \csname LTb\endcsname
      \put(462,1941){\rotatebox{-270}{\makebox(0,0){\strut{}Intensity, a.u.}}}%
      \put(2825,154){\makebox(0,0){\strut{}Photon energy $\omega$, eV}}%
      \csname LTb\endcsname
      \put(1933,3006){\makebox(0,0)[l]{\strut{}THERMOS\_CRE}}%
      \csname LTb\endcsname
      \put(1933,2786){\makebox(0,0)[l]{\strut{}experiment}}%
    }%
    \gplbacktext
    \put(0,0){\includegraphics{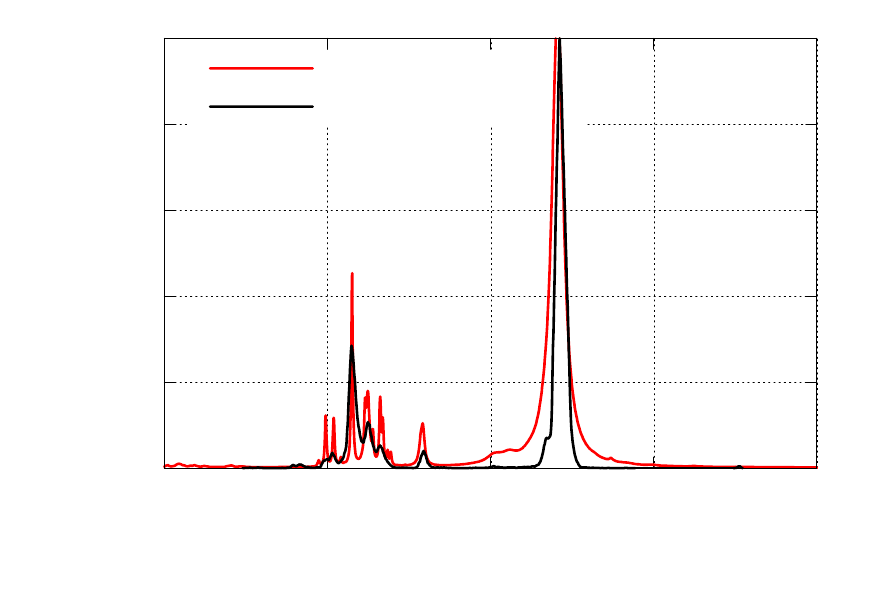}}%
    \gplfronttext
  \end{picture}%
\endgroup
\caption{Calculated emission spectrum of the chlorine plasma at the electron density $10^{23}$~cm$^{-3}$ and temperature~600 eV in comparison with the ORION experiment \cite{doi:10.1063/1.4965233}. The layer thickness for the THERMOS\_CRE calculation was set equal to $L= 4\times 10^{-5}$~cm.}
\label{Cl_exp}
\end{figure}

In the experimental emission spectrum of the chlorine plasma, shown in Fig.~\ref{Cl_exp}, two groups of lines can be distinguished. The first one corresponds to the $1s3p\rightarrow 1s^2$ transitions in the helium-like Cl$^{+15}$ ion. The second group corresponds to the lithium-like Cl$^{+14}$ ion transitions $1s2s3p \rightarrow 1s^2 2s$ and $1s2p3p \rightarrow 1s^2 2p$. The best fit solution was found by adjusting the ratio of the intensities of the Cl$^{+14}$ satellite lines $I(1s2s3p \rightarrow 1s^2 2s)/I(1s2p3p \rightarrow 1s^2 2p)$, calculated by solving the system of the level kinetics equations, to its measured value. In result, the best agreement was achieved for the electron density $N_e= 10^{23}$~cm$^{-3}$ and the temperature $T_e= 600$~eV.

\begin{figure}[!htb]
 \centering
\begingroup
  \makeatletter
  \providecommand\color[2][]{%
    \GenericError{(gnuplot) \space\space\space\@spaces}{%
      Package color not loaded in conjunction with
      terminal option `colourtext'%
    }{See the gnuplot documentation for explanation.%
    }{Either use 'blacktext' in gnuplot or load the package
      color.sty in LaTeX.}%
    \renewcommand\color[2][]{}%
  }%
  \providecommand\includegraphics[2][]{%
    \GenericError{(gnuplot) \space\space\space\@spaces}{%
      Package graphicx or graphics not loaded%
    }{See the gnuplot documentation for explanation.%
    }{The gnuplot epslatex terminal needs graphicx.sty or graphics.sty.}%
    \renewcommand\includegraphics[2][]{}%
  }%
  \providecommand\rotatebox[2]{#2}%
  \@ifundefined{ifGPcolor}{%
    \newif\ifGPcolor
    \GPcolortrue
  }{}%
  \@ifundefined{ifGPblacktext}{%
    \newif\ifGPblacktext
    \GPblacktexttrue
  }{}%
  \let\gplgaddtomacro\g@addto@macro
  \gdef\gplbacktext{}%
  \gdef\gplfronttext{}%
  \makeatother
  \ifGPblacktext
    \def\colorrgb#1{}%
    \def\colorgray#1{}%
  \else
    \ifGPcolor
      \def\colorrgb#1{\color[rgb]{#1}}%
      \def\colorgray#1{\color[gray]{#1}}%
      \expandafter\def\csname LTw\endcsname{\color{white}}%
      \expandafter\def\csname LTb\endcsname{\color{black}}%
      \expandafter\def\csname LTa\endcsname{\color{black}}%
      \expandafter\def\csname LT0\endcsname{\color[rgb]{1,0,0}}%
      \expandafter\def\csname LT1\endcsname{\color[rgb]{0,1,0}}%
      \expandafter\def\csname LT2\endcsname{\color[rgb]{0,0,1}}%
      \expandafter\def\csname LT3\endcsname{\color[rgb]{1,0,1}}%
      \expandafter\def\csname LT4\endcsname{\color[rgb]{0,1,1}}%
      \expandafter\def\csname LT5\endcsname{\color[rgb]{1,1,0}}%
      \expandafter\def\csname LT6\endcsname{\color[rgb]{0,0,0}}%
      \expandafter\def\csname LT7\endcsname{\color[rgb]{1,0.3,0}}%
      \expandafter\def\csname LT8\endcsname{\color[rgb]{0.5,0.5,0.5}}%
    \else
      \def\colorrgb#1{\color{black}}%
      \def\colorgray#1{\color[gray]{#1}}%
      \expandafter\def\csname LTw\endcsname{\color{white}}%
      \expandafter\def\csname LTb\endcsname{\color{black}}%
      \expandafter\def\csname LTa\endcsname{\color{black}}%
      \expandafter\def\csname LT0\endcsname{\color{black}}%
      \expandafter\def\csname LT1\endcsname{\color{black}}%
      \expandafter\def\csname LT2\endcsname{\color{black}}%
      \expandafter\def\csname LT3\endcsname{\color{black}}%
      \expandafter\def\csname LT4\endcsname{\color{black}}%
      \expandafter\def\csname LT5\endcsname{\color{black}}%
      \expandafter\def\csname LT6\endcsname{\color{black}}%
      \expandafter\def\csname LT7\endcsname{\color{black}}%
      \expandafter\def\csname LT8\endcsname{\color{black}}%
    \fi
  \fi
    \setlength{\unitlength}{0.0500bp}%
    \ifx\gptboxheight\undefined%
      \newlength{\gptboxheight}%
      \newlength{\gptboxwidth}%
      \newsavebox{\gptboxtext}%
    \fi%
    \setlength{\fboxrule}{0.5pt}%
    \setlength{\fboxsep}{1pt}%
\begin{picture}(5102.00,3400.00)%
    \gplgaddtomacro\gplbacktext{%
      \csname LTb\endcsname
      \put(814,704){\makebox(0,0)[r]{\strut{}$0.5$}}%
      \csname LTb\endcsname
      \put(814,1199){\makebox(0,0)[r]{\strut{}$0.6$}}%
      \csname LTb\endcsname
      \put(814,1694){\makebox(0,0)[r]{\strut{}$0.7$}}%
      \csname LTb\endcsname
      \put(814,2189){\makebox(0,0)[r]{\strut{}$0.8$}}%
      \csname LTb\endcsname
      \put(814,2684){\makebox(0,0)[r]{\strut{}$0.9$}}%
      \csname LTb\endcsname
      \put(814,3179){\makebox(0,0)[r]{\strut{}$1$}}%
      \csname LTb\endcsname
      \put(946,484){\makebox(0,0){\strut{}$3150$}}%
      \csname LTb\endcsname
      \put(1886,484){\makebox(0,0){\strut{}$3200$}}%
      \csname LTb\endcsname
      \put(2826,484){\makebox(0,0){\strut{}$3250$}}%
      \csname LTb\endcsname
      \put(3765,484){\makebox(0,0){\strut{}$3300$}}%
      \csname LTb\endcsname
      \put(4705,484){\makebox(0,0){\strut{}$3350$}}%
    }%
    \gplgaddtomacro\gplfronttext{%
      \csname LTb\endcsname
      \put(462,1941){\rotatebox{-270}{\makebox(0,0){\strut{}Transmission}}}%
      \put(2825,154){\makebox(0,0){\strut{}Photon energy $\omega$, eV}}%
      \csname LTb\endcsname
      \put(1933,1097){\makebox(0,0)[l]{\strut{}$4\cdot 10^{-5}$ cm}}%
      \csname LTb\endcsname
      \put(1933,877){\makebox(0,0)[l]{\strut{}10$^{-4}$ cm}}%
    }%
    \gplbacktext
    \put(0,0){\includegraphics{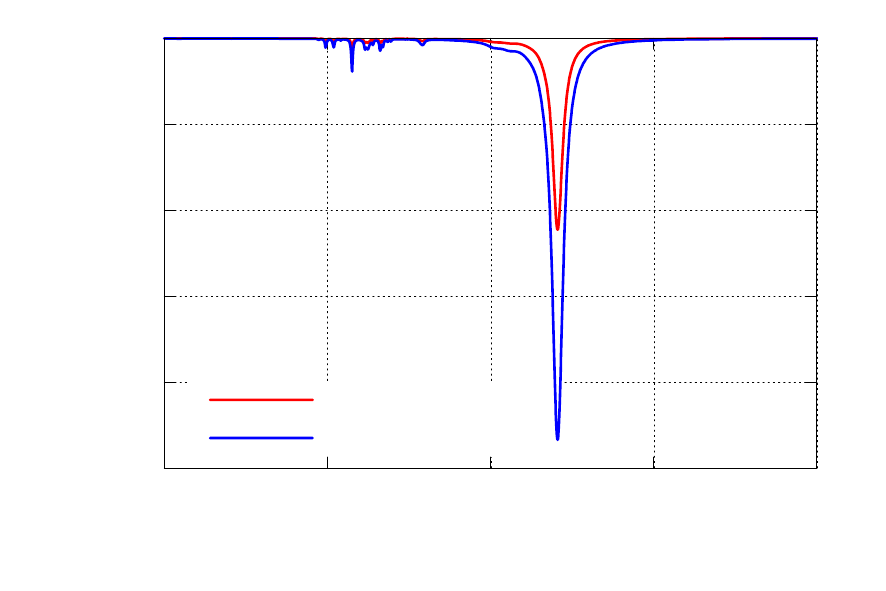}}%
    \gplfronttext
  \end{picture}%
\endgroup
\caption{Transmission spectra of the chlorine plasma at the electron density $10^{23}$~cm$^{-3}$ and the temperature~600 eV for the layer thicknesses $L= 4\times 10^{-5}$ and 10$^{-4}$~cm.}
\label{Cl:trans}
\end{figure}

Under these conditions the chlorine plasma is optically thick, so one can assume that the measured radiation originates from a thin surface layer. Figure~\ref{Cl:trans} shows two transmission spectra of the chlorine plasma at $N_e= 10^{23}$~cm$^{-3}$ and $T_e= 600$~eV, calculated for two values of the layer thickness $L$, namely, for $L= 4\times 10^{-5}$ and 10$^{-4}$~cm. The transmission of a layer of thickness $L$ at a photon energy $\omega$ was calculated as
\begin{equation}
 T_{\omega} =\exp(-\varkappa_{\omega}\rho L),
\label{transL}
\end{equation}
where $\varkappa_{\omega}$ (cm$^2$/g) is the opacity, and $\rho$ is the mass density. Figure~\ref{Cl:trans} clearly indicates that for $L= 4\times 10^{-5}$~cm there already is a significant amount of reabsorption in the $1s3p \rightarrow 1s^2$ line, while for greater thicknesses $L \geq 10^{-4}$~cm the reabsorption must be taken into account for practically all the observed spectral lines.

The radiation flux, emitted by a uniform slab with a thickness $L$ at a photon energy $\omega$, can be calculated as
\begin{equation}
F_{\omega} = \pi \frac{j_{\omega}}{\rho\varkappa_{\omega}}\left[1-e^{-\tau} +\tau e^{-\tau} -\tau^2 E_1(\tau)\right],
\label{fluxL}
\end{equation}
where $E_k(\tau) =\int\limits_{\tau}^{\infty} t^{-k}e^{-t}dt$, $\tau= L\rho \varkappa_{\omega}$ is the optical thickness, and $j_{\omega}$ and $\varkappa_{\omega}$ are, respectively, the spectral emissivity and the opacity. The radiation flux, calculated in this way for $L= 4\times 10^{-5}$~cm with the $j_{\omega}$ and $\varkappa_{\omega}$ provided by the THERMOS\_CRE code for $N_e= 10^{23}$~cm$^{-3}$ and $T_e= $600~eV, is shown in Fig.~\ref{Cl_exp} together with the experimental data. It is seen that a fair agreement with the measured emission spectrum can be achieved by using the approximation of a thin uniform layer.

\section{Transmission of a silicon plasma slab}
\label{subSection_Si_slab}

At the 10$^{th}$ NLTE Workshop, a special interest was paid to the results of Ref.~\cite{PhysRevLett.119.075001}, where the experimental transmission spectrum for soft X-rays through a slab of photoionized silicon plasma was presented --- under the conditions that are close to astrophysical environment. In such a plasma, radiative processes dominate over the collisional ones due to the presence of an intense external radiation field, which pumps the autoionization states and strongly enhances the role of the Auger transitions.

For the numerical modeling of this experiment, it was proposed to consider a homogeneous slab of a silicon plasma, whose thickness $L$, temperature $T_e$, and the electron density $N_e$ are treated as free parameters; for the intensity $I_{\omega}$ of the ambient radiation field, it was proposed to use the diluted black-body spectrum
\begin{equation}
I_{\omega} =B_{\omega} = K \frac{15 \sigma}{\pi^5}
\frac{\omega^3}{\exp(\omega /T_{rad}) - 1},
\label{rad_W_omega}
\end{equation}
where $K$ is the dilution factor, $\omega$ is the photon energy, $T_{rad}$ is the radiation temperature, and $\sigma$ is the Stefan-Boltzmann constant; $I_{\omega}$ and $B_{\omega}$ are defined per unit photon energy interval. The plasma parameters were supposed to be determined by the best fit to the measured transmission spectrum. The analysis, carried out in Ref.~\cite{PhysRevLett.119.075001}, allowed to outline the bounds for the search area: the plasma size should be in the range $L \simeq 0.1$--1~cm, the electron density $N_e \simeq 10^{19}$~cm$^{-3}$, and the temperature $T_e\simeq 30$~eV, with the mean ion charge being around $Z_0\approx 10$.

First we consider one of the cases proposed for the preliminary analysis, namely, $L= 0.1$~cm, $N_e =10^{19}$~cm$^{-3}$, $T_e= 30$~eV, $K=1$, and $T_{rad}=63$~eV. For further discussion, this case is denoted the ``uniform slab'', which means that the electron temperature and density, as well as the radiation energy density  $\displaystyle U_{\omega}=\frac{4\pi}{c}B_{\omega}$ are constant over the plasma volume. Calculation using the THERMOS\_CRE code for a uniform slab with the radiation field~(\ref{rad_W_omega}) resulted in the mean charge $Z_0=11.8$. This value does not match the estimated optimum of $Z_0~\approx$~10, and, accordingly, the calculated transmission spectrum (the blue curve in Fig.~\ref{Si_exp}) is rather far from the experimental one. In the course of discussions at the Workshop, it was decided to diminish the role of the ambient radiation field by setting $K=0.1$, which led to the mean ion charge lying closer to 10 and to a better agreement with the experimental transmission spectrum. Here, the dilution factor of 0.1 made it possible to artificially take into account the effects of the reabsorption of radiation in a dense plasma.

At the same time, a more thorough analysis of the opacity of the silicon plasma revealed that the major part of the external radiation flux was absorbed in a thin edge layer, and was unable to penetrate deep into the slab. For the relevant values of the temperature, electron density and radiation field, a layer with a thickness in excess of $10^{-3}$~cm is opaque for the photons in the investigated wavelength range (6.64--6.96~\AA). Hence, the spatial distribution of the radiation field must be treated as non-uniform. In order to take into account the effect of reabsorption in the plasma slab, we have modified the problem statement and the numerical model.

\begin{figure}[!htb]
	\centering
    \includegraphics[width=0.9\linewidth]{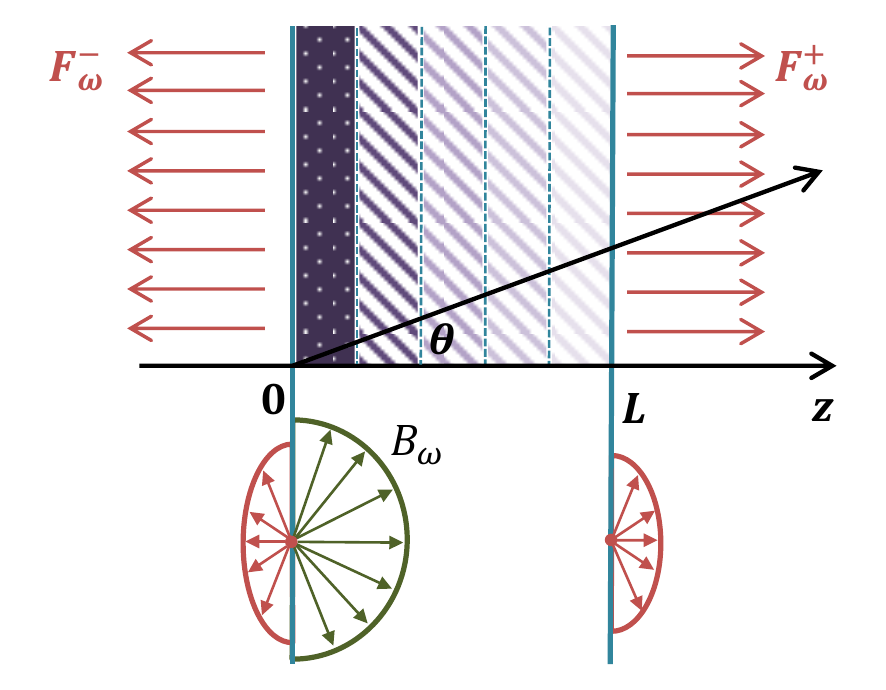}
	\caption{Schematic formulation of the problem for the silicon plasma.}
	\label{Layer_Si}
\end{figure}

In the new formulation, the initial slab with $L=0.1$~cm is divided into $N$ layers ($z_{i-1}<z<z_{i},\, i=1,N$). In each layer, the electron density and temperature are kept the same as in the initial case of a uniform slab, i.e.\ $N_e= 10^{19}$~cm$^{-3}$ and $T_e= 30$~eV. Also, the radiation energy density $U_{\omega}=\displaystyle\frac{1}{c}\int_{4\pi} I_{\omega} d\Omega$ is assumed to be constant within each layer, but, in contrast to $N_e$ and $T_e$, it is allowed to vary from layer to layer. This case is called the ``multi-layer slab'' (see Fig.~\ref{Layer_Si}).

The radiation intensity $I_{\omega}(z,\theta)$ is found as a solution of the transfer equation
\begin{equation}
 \mu \frac{dI_{\omega}}{dz} = j_{\omega}-\rho\varkappa_{\omega}I_{\omega},
 \label{rad_L}
 \end{equation}
where $\mu=\cos \theta$. The emissivity $j_{\omega}(z, n_{js})$ and the opacity $\varkappa_{\omega}(z, n_{js})$ at given $T_e$, $N_e$, and $U_{\omega}$ are determined by the populations $n_{js}(z)$ of the ion species with the ionization degree~$j$ in a quantum state~$s$. The values of $n_{js}(z)$ are found as a solution of the system of rate equations
\begin{equation}
 \sum\limits_{j's'}\left( n_{j's'}w_{j's'\to js} - n_{js} w_{js\to j's'} \right) = 0,
 \label{CRE}
 \end{equation}
where $w_{js \to j's'}(U_{\omega})$ is the total probability of transition from an ion state $js$ to a state $j's'$ due to the relevant elementary processes~\cite{Nikiforov2005}.

To calculate the plasma radiative properties, one has to solve self-consistently the transfer equation (\ref{rad_L}) together with the rate equations (\ref{CRE}). The boundary condition for Eq.~(\ref{rad_L}) takes into account the incident flux of the external radiation,
\begin{equation}
 I_{\omega}(z=0, \mu >0)=B_{\omega},
 \label{rad_Bl}
\end{equation}
\begin{equation}
 I_{\omega}(z=L, \mu <0)=0,
 \label{rad_B}
\end{equation}
where $B_{\omega}$ is given by Eq.~(\ref{rad_W_omega}) with $K=1$ and $T_{rad} =$63~eV. The self-consistent solution of equations~(\ref{rad_L}) and (\ref{CRE}) for $I_{\omega}(z)$ and $n_{js}(z)$ is obtained by successive iterations. At each iteration, the transfer equation (\ref{rad_L}) is solved exactly for given piece-wise constant coefficients $j_{\omega}(z)$ and $\rho\varkappa_{\omega}(z)$ --- in accordance with the discretization of $z$.  At the initial iteration $\nu=0$, the level populations in the first layer $z_0< z <z_1$ are calculated with the $\displaystyle U_{\omega}= \frac{4\pi}{c}B_{\omega}$, and in the remaining $N-1$ layers at $z_1< z< z_N$  with the $U_{\omega}=0$. All the subsequent iterations $\nu \geq 1$ use the values $U_{\omega}^{(\nu-1)}$, calculated at the previous step $\nu-1$. The iterations are assumed to have converged when the relative accuracy of $10^{-3}$ is reached in the $L_1$ norm. The radiation flux at the right boundary is calculated as
\begin{equation}
F_{\omega}^{+} =2\pi \int\limits_0^{\frac{\pi}{2}}I_{\omega}(z=L) \cos \theta \sin \theta d \theta.
\label{rad_F}
\end{equation}

\begin{figure}[!htb]
 \centering
\begingroup
  \makeatletter
  \providecommand\color[2][]{%
    \GenericError{(gnuplot) \space\space\space\@spaces}{%
      Package color not loaded in conjunction with
      terminal option `colourtext'%
    }{See the gnuplot documentation for explanation.%
    }{Either use 'blacktext' in gnuplot or load the package
      color.sty in LaTeX.}%
    \renewcommand\color[2][]{}%
  }%
  \providecommand\includegraphics[2][]{%
    \GenericError{(gnuplot) \space\space\space\@spaces}{%
      Package graphicx or graphics not loaded%
    }{See the gnuplot documentation for explanation.%
    }{The gnuplot epslatex terminal needs graphicx.sty or graphics.sty.}%
    \renewcommand\includegraphics[2][]{}%
  }%
  \providecommand\rotatebox[2]{#2}%
  \@ifundefined{ifGPcolor}{%
    \newif\ifGPcolor
    \GPcolortrue
  }{}%
  \@ifundefined{ifGPblacktext}{%
    \newif\ifGPblacktext
    \GPblacktexttrue
  }{}%
  \let\gplgaddtomacro\g@addto@macro
  \gdef\gplbacktext{}%
  \gdef\gplfronttext{}%
  \makeatother
  \ifGPblacktext
    \def\colorrgb#1{}%
    \def\colorgray#1{}%
  \else
    \ifGPcolor
      \def\colorrgb#1{\color[rgb]{#1}}%
      \def\colorgray#1{\color[gray]{#1}}%
      \expandafter\def\csname LTw\endcsname{\color{white}}%
      \expandafter\def\csname LTb\endcsname{\color{black}}%
      \expandafter\def\csname LTa\endcsname{\color{black}}%
      \expandafter\def\csname LT0\endcsname{\color[rgb]{1,0,0}}%
      \expandafter\def\csname LT1\endcsname{\color[rgb]{0,1,0}}%
      \expandafter\def\csname LT2\endcsname{\color[rgb]{0,0,1}}%
      \expandafter\def\csname LT3\endcsname{\color[rgb]{1,0,1}}%
      \expandafter\def\csname LT4\endcsname{\color[rgb]{0,1,1}}%
      \expandafter\def\csname LT5\endcsname{\color[rgb]{1,1,0}}%
      \expandafter\def\csname LT6\endcsname{\color[rgb]{0,0,0}}%
      \expandafter\def\csname LT7\endcsname{\color[rgb]{1,0.3,0}}%
      \expandafter\def\csname LT8\endcsname{\color[rgb]{0.5,0.5,0.5}}%
    \else
      \def\colorrgb#1{\color{black}}%
      \def\colorgray#1{\color[gray]{#1}}%
      \expandafter\def\csname LTw\endcsname{\color{white}}%
      \expandafter\def\csname LTb\endcsname{\color{black}}%
      \expandafter\def\csname LTa\endcsname{\color{black}}%
      \expandafter\def\csname LT0\endcsname{\color{black}}%
      \expandafter\def\csname LT1\endcsname{\color{black}}%
      \expandafter\def\csname LT2\endcsname{\color{black}}%
      \expandafter\def\csname LT3\endcsname{\color{black}}%
      \expandafter\def\csname LT4\endcsname{\color{black}}%
      \expandafter\def\csname LT5\endcsname{\color{black}}%
      \expandafter\def\csname LT6\endcsname{\color{black}}%
      \expandafter\def\csname LT7\endcsname{\color{black}}%
      \expandafter\def\csname LT8\endcsname{\color{black}}%
    \fi
  \fi
  \setlength{\unitlength}{0.0500bp}%
  \begin{picture}(5102.00,3400.00)%
    \gplgaddtomacro\gplbacktext{%
      \csname LTb\endcsname%
      \put(946,1190){\makebox(0,0)[r]{\strut{}10$^{-6}$}}%
      \csname LTb\endcsname%
      \put(946,1676){\makebox(0,0)[r]{\strut{}10$^{-4}$}}%
      \csname LTb\endcsname%
      \put(946,2163){\makebox(0,0)[r]{\strut{}10$^{-2}$}}%
      \csname LTb\endcsname%
      \put(946,2649){\makebox(0,0)[r]{\strut{}1}}%
      \csname LTb\endcsname%
      \put(946,3135){\makebox(0,0)[r]{\strut{}10$^{2}$}}%
      \csname LTb\endcsname%
      \put(946,704){\makebox(0,0)[r]{\strut{}}}%
      \csname LTb\endcsname%
      \put(946,947){\makebox(0,0)[r]{\strut{}}}%
      \csname LTb\endcsname%
      \put(946,1433){\makebox(0,0)[r]{\strut{}}}%
      \csname LTb\endcsname%
      \put(946,1920){\makebox(0,0)[r]{\strut{}}}%
      \csname LTb\endcsname%
      \put(946,2406){\makebox(0,0)[r]{\strut{}}}%
      \csname LTb\endcsname%
      \put(946,2892){\makebox(0,0)[r]{\strut{}}}%
      \csname LTb\endcsname%
      \put(1078,484){\makebox(0,0){\strut{} 1700}}%
      \csname LTb\endcsname%
      \put(1985,484){\makebox(0,0){\strut{} 1900}}%
      \csname LTb\endcsname%
      \put(2892,484){\makebox(0,0){\strut{} 2100}}%
      \csname LTb\endcsname%
      \put(3798,484){\makebox(0,0){\strut{} 2300}}%
      \csname LTb\endcsname%
      \put(4705,484){\makebox(0,0){\strut{} 2500}}%
      \put(176,1919){\rotatebox{-270}{\makebox(0,0){\strut{}Radiation flux $F_{\omega}^+$, J/s/cm$^2$/eV}}}%
      \put(2891,154){\makebox(0,0){\strut{}Photon energy $\omega$, eV}}%
    }%
    \gplgaddtomacro\gplfronttext{%
      \csname LTb\endcsname%
      \put(3718,2962){\makebox(0,0)[r]{\strut{}uniform slab}}%
      \csname LTb\endcsname%
      \put(3718,2742){\makebox(0,0)[r]{\strut{}multi-layer slab}}%
    }%
    \gplbacktext
    \put(0,0){\includegraphics{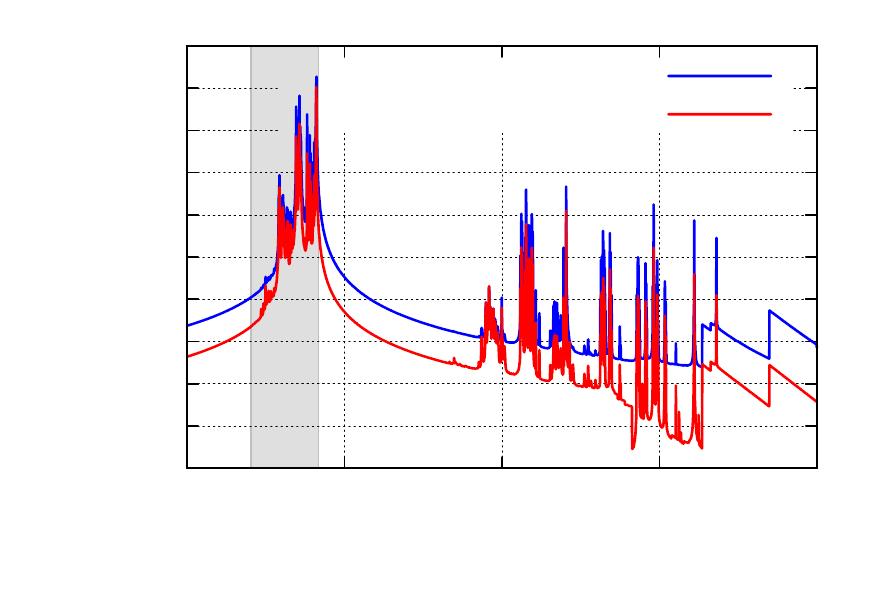}}%
    \gplfronttext
  \end{picture}%
\endgroup
	\caption{Radiation flux at the right boundary of the silicon plasma slab with $L=0.1$~cm for the ``uniform slab'' and the ``multi-layer slab'' cases. The gray vertical band represents the spectral interval investigated in Ref.~\cite{PhysRevLett.119.075001} and displayed in Fig.~\ref{Si_exp}.}
	\label{Si_2models}
\end{figure}

\begin{figure}[!htb]
 \centering
\begingroup
  \makeatletter
  \providecommand\color[2][]{%
    \GenericError{(gnuplot) \space\space\space\@spaces}{%
      Package color not loaded in conjunction with
      terminal option `colourtext'%
    }{See the gnuplot documentation for explanation.%
    }{Either use 'blacktext' in gnuplot or load the package
      color.sty in LaTeX.}%
    \renewcommand\color[2][]{}%
  }%
  \providecommand\includegraphics[2][]{%
    \GenericError{(gnuplot) \space\space\space\@spaces}{%
      Package graphicx or graphics not loaded%
    }{See the gnuplot documentation for explanation.%
    }{The gnuplot epslatex terminal needs graphicx.sty or graphics.sty.}%
    \renewcommand\includegraphics[2][]{}%
  }%
  \providecommand\rotatebox[2]{#2}%
  \@ifundefined{ifGPcolor}{%
    \newif\ifGPcolor
    \GPcolortrue
  }{}%
  \@ifundefined{ifGPblacktext}{%
    \newif\ifGPblacktext
    \GPblacktexttrue
  }{}%
  \let\gplgaddtomacro\g@addto@macro
  \gdef\gplbacktext{}%
  \gdef\gplfronttext{}%
  \makeatother
  \ifGPblacktext
    \def\colorrgb#1{}%
    \def\colorgray#1{}%
  \else
    \ifGPcolor
      \def\colorrgb#1{\color[rgb]{#1}}%
      \def\colorgray#1{\color[gray]{#1}}%
      \expandafter\def\csname LTw\endcsname{\color{white}}%
      \expandafter\def\csname LTb\endcsname{\color{black}}%
      \expandafter\def\csname LTa\endcsname{\color{black}}%
      \expandafter\def\csname LT0\endcsname{\color[rgb]{1,0,0}}%
      \expandafter\def\csname LT1\endcsname{\color[rgb]{0,1,0}}%
      \expandafter\def\csname LT2\endcsname{\color[rgb]{0,0,1}}%
      \expandafter\def\csname LT3\endcsname{\color[rgb]{1,0,1}}%
      \expandafter\def\csname LT4\endcsname{\color[rgb]{0,1,1}}%
      \expandafter\def\csname LT5\endcsname{\color[rgb]{1,1,0}}%
      \expandafter\def\csname LT6\endcsname{\color[rgb]{0,0,0}}%
      \expandafter\def\csname LT7\endcsname{\color[rgb]{1,0.3,0}}%
      \expandafter\def\csname LT8\endcsname{\color[rgb]{0.5,0.5,0.5}}%
    \else
      \def\colorrgb#1{\color{black}}%
      \def\colorgray#1{\color[gray]{#1}}%
      \expandafter\def\csname LTw\endcsname{\color{white}}%
      \expandafter\def\csname LTb\endcsname{\color{black}}%
      \expandafter\def\csname LTa\endcsname{\color{black}}%
      \expandafter\def\csname LT0\endcsname{\color{black}}%
      \expandafter\def\csname LT1\endcsname{\color{black}}%
      \expandafter\def\csname LT2\endcsname{\color{black}}%
      \expandafter\def\csname LT3\endcsname{\color{black}}%
      \expandafter\def\csname LT4\endcsname{\color{black}}%
      \expandafter\def\csname LT5\endcsname{\color{black}}%
      \expandafter\def\csname LT6\endcsname{\color{black}}%
      \expandafter\def\csname LT7\endcsname{\color{black}}%
      \expandafter\def\csname LT8\endcsname{\color{black}}%
    \fi
  \fi
  \setlength{\unitlength}{0.0500bp}%
  \begin{picture}(5102.00,3400.00)%
    \gplgaddtomacro\gplbacktext{%
      \csname LTb\endcsname%
      \put(726,704){\makebox(0,0)[r]{\strut{} 0}}%
      \csname LTb\endcsname%
      \put(726,1190){\makebox(0,0)[r]{\strut{} 0.2}}%
      \csname LTb\endcsname%
      \put(726,1676){\makebox(0,0)[r]{\strut{} 0.4}}%
      \csname LTb\endcsname%
      \put(726,2163){\makebox(0,0)[r]{\strut{} 0.6}}%
      \csname LTb\endcsname%
      \put(726,2649){\makebox(0,0)[r]{\strut{} 0.8}}%
      \csname LTb\endcsname%
      \put(726,3135){\makebox(0,0)[r]{\strut{} 1}}%
      \csname LTb\endcsname%
      \put(978,484){\makebox(0,0){\strut{} 6.65}}%
      \csname LTb\endcsname%
      \put(1579,484){\makebox(0,0){\strut{} 6.7}}%
      \csname LTb\endcsname%
      \put(2180,484){\makebox(0,0){\strut{} 6.75}}%
      \csname LTb\endcsname%
      \put(2782,484){\makebox(0,0){\strut{} 6.8}}%
      \csname LTb\endcsname%
      \put(3383,484){\makebox(0,0){\strut{} 6.85}}%
      \csname LTb\endcsname%
      \put(3984,484){\makebox(0,0){\strut{} 6.9}}%
      \csname LTb\endcsname%
      \put(4585,484){\makebox(0,0){\strut{} 6.95}}%
      \put(220,1919){\rotatebox{-270}{\makebox(0,0){\strut{}Transmission}}}%
      \put(2781,154){\makebox(0,0){\strut{}Wavelength, \AA}}%
    }%
    \gplgaddtomacro\gplfronttext{%
      \csname LTb\endcsname%
      \put(3718,1317){\makebox(0,0)[r]{\strut{}experiment}}%
      \csname LTb\endcsname%
      \put(3718,1097){\makebox(0,0)[r]{\strut{}uniform slab}}%
      \csname LTb\endcsname%
      \put(3718,877){\makebox(0,0)[r]{\strut{}multi-layer slab}}%
    }%
    \gplbacktext
    \put(0,0){\includegraphics{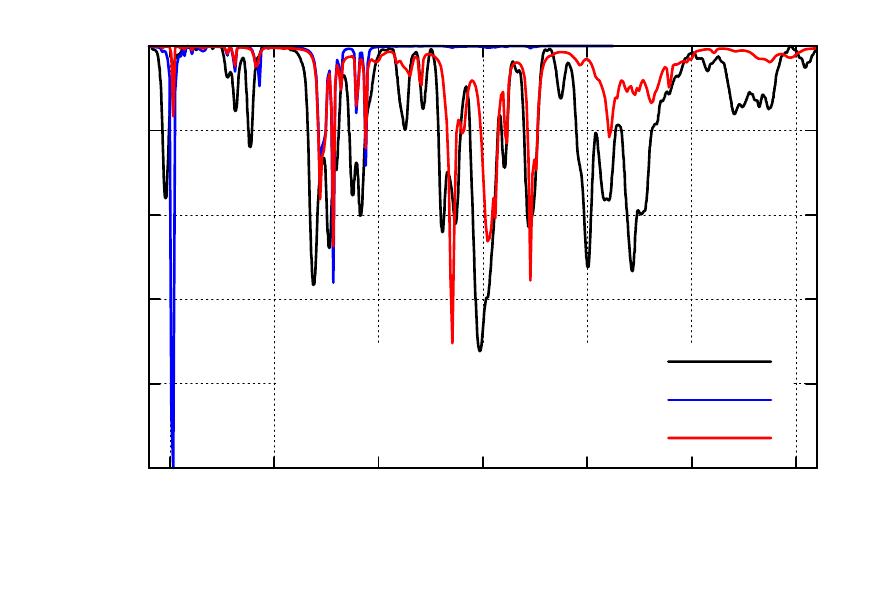}}%
    \gplfronttext
  \end{picture}%
\endgroup
	\caption{Two calculated transmission spectra of a silicon plasma slab with the thickness $L=0.1$~cm, the electron density 10$^{19}$~cm$^{-3}$ and the temperature 30~eV compared to the experimental data from Ref.~\cite{PhysRevLett.119.075001}.
	}
	\label{Si_exp}
\end{figure}

The reference calculation, compared with the experiment, was performed for a slab divided into $N=5$ layers. All the layer thicknesses were equal to $L/N$. The obtained solution has a non-uniform distribution of the radiation intensity in the slab. Figure~\ref{Si_2models} compares the two spectral fluxes, emitted from the right boundary $z=L$, as calculated from Eq.~(\ref{rad_F}) in the multi-layer approximation (the red curve), and for the uniform slab from Eq.~(\ref{fluxL}) (the blue curve). The two spectra significantly differ from each other because in the ``multi-layer slab'' case the mean ion charge decreases from $Z_0=11.6$ in the leftmost layer to $Z_0=9.7$ in the rightmost layer, whereas for the ``uniform slab'' it does not depend on $z$ and is everywhere equal to $Z_0=11.8$.

Figure~\ref{Si_exp} shows the transmission spectra for the considered silicon plasma slab: the two theoretical spectra, calculated respectively in the multi-layer approximation (the red curve) and for the uniform slab (the blue curve), are compared with the measured spectrum (the black curve) from Ref.~\cite{PhysRevLett.119.075001}. One clearly sees that the ``multi-layer slab'' spectrum gives a significantly better agreement with the experiment than the transmission of a uniform slab because the non-uniformity of the radiation field along the $z$-axis is properly accounted for. The deviations in the positions and strengths of individual lines are explained by the use of non-relativistic configurations when solving the rate equations (\ref{CRE}).

\section{Summary}
Some of the key features of the THERMOS toolkit for modeling radiative plasma properties are described. The impact of the plasma density effects is analyzed for the aluminum and chlorine plasma cases. The THERMOS simulation results demonstrate a reasonable agreement with the experimental data, as well as with the other codes that participated in the 10$^{th}$ NLTE Workshop. The model of self-consistent calculation of radiation transfer and level kinetics in a multi-layer plasma slab was used to account for the radiation reabsorption, important for the case of non-equilibrium silicon plasma. In this way, a good agreement with the experimental transmission spectrum was achieved, except for the line strengths of certain ion species.

\section*{Acknowledgment}
Authors thank the participants and organizers of the NLTE and RPHDM conferences for productive discussions of the problems arising from non-equilibrium plasma simulations. This work was funded by the Russian Science Foundation through grant No.~14-11-00699. Calculations have been performed at HPC MVS-10P (JSCC RAS) and HPC K100 (KIAM RAS).


\bibliography{elsarticle-rphdm-vichev}

\begin{thebibliography}{10}
\expandafter\ifx\csname url\endcsname\relax
  \def\url#1{\texttt{#1}}\fi
\expandafter\ifx\csname urlprefix\endcsname\relax\def\urlprefix{URL }\fi
\expandafter\ifx\csname href\endcsname\relax
  \def\href#1#2{#2} \def\path#1{#1}\fi

\bibitem{doi:10.1063/1.4965233}
P.~Beiersdorfer, G.~V. Brown, R.~Shepherd, P.~Allan, C.~R.~D. Brown, M.~P.
  Hill, D.~J. Hoarty, L.~M.~R. Hobbs, S.~F. James, H.~K. Chung, E.~Hill,
  \href{https://doi.org/10.1063/1.4965233}{Lineshape measurements of he-$\beta$
  spectra on the orion laser facility}, Physics of Plasmas 23~(10) (2016)
  101211.
\newblock \href {http://arxiv.org/abs/https://doi.org/10.1063/1.4965233}
  {\path{arXiv:https://doi.org/10.1063/1.4965233}}, \href
  {https://doi.org/10.1063/1.4965233} {\path{doi:10.1063/1.4965233}}.
\newline\urlprefix\url{https://doi.org/10.1063/1.4965233}

\bibitem{PhysRevLett.119.075001}
G.~P. Loisel, J.~E. Bailey, D.~A. Liedahl, C.~J. Fontes, T.~R. Kallman,
  T.~Nagayama, S.~B. Hansen, G.~A. Rochau, R.~C. Mancini, R.~W. Lee,
  \href{https://link.aps.org/doi/10.1103/PhysRevLett.119.075001}{Benchmark
  experiment for photoionized plasma emission from accretion-powered x-ray
  sources}, Phys. Rev. Lett. 119 (2017) 075001.
\newblock \href {https://doi.org/10.1103/PhysRevLett.119.075001}
  {\path{doi:10.1103/PhysRevLett.119.075001}}.
\newline\urlprefix\url{https://link.aps.org/doi/10.1103/PhysRevLett.119.075001}

\bibitem{Nikiforov2005}
A.~F. Nikiforov, V.~G. Novikov, V.~B. Uvarov, {Quantum-statistical models of
  hot dense matter. Methods for computation opacity and equation of state.},
  Birkh{\"{a}}user., Switzerland, 2005.

\bibitem{FAIK201447}
S.~Faik, A.~Tauschwitz, M.~M. Basko, J.~A. Maruhn, O.~Rosmej, T.~Rienecker,
  V.~G. Novikov, A.~S. Grushin,
  \href{http://www.sciencedirect.com/science/article/pii/S1574181813001730}{Creation
  of a homogeneous plasma column by means of hohlraum radiation for
  ion-stopping measurements}, High Energy Density Physics 10 (2014) 47 -- 55.
\newblock \href {https://doi.org/https://doi.org/10.1016/j.hedp.2013.10.002}
  {\path{doi:https://doi.org/10.1016/j.hedp.2013.10.002}}.
\newline\urlprefix\url{http://www.sciencedirect.com/science/article/pii/S1574181813001730}

\bibitem{1742-6596-653-1-012148}
O.~G. Olkhovskaya, M.~M. Basko, P.~V. Sasorov, I.~Y. Vitchev, V.~G. Novikov,
  A.~S. Boldarev, V.~A. Gasilov, S.~I. Tkachenko,
  \href{http://stacks.iop.org/1742-6596/653/i=1/a=012148}{Radiative power and
  x-ray spectrum numerical estimations for wire array z-pinches}, Journal of
  Physics: Conference Series 653~(1) (2015) 012148.
\newline\urlprefix\url{http://stacks.iop.org/1742-6596/653/i=1/a=012148}

\bibitem{doi:10.1063/1.4921334}
M.~M. Basko, V.~G. Novikov, A.~S. Grushin,
  \href{https://aip.scitation.org/doi/abs/10.1063/1.4921334}{On the structure
  of quasi-stationary laser ablation fronts in strongly radiating plasmas},
  Physics of Plasmas 22~(5) (2015) 053111.
\newblock \href
  {http://arxiv.org/abs/https://aip.scitation.org/doi/pdf/10.1063/1.4921334}
  {\path{arXiv:https://aip.scitation.org/doi/pdf/10.1063/1.4921334}}, \href
  {https://doi.org/10.1063/1.4921334} {\path{doi:10.1063/1.4921334}}.
\newline\urlprefix\url{https://aip.scitation.org/doi/abs/10.1063/1.4921334}

\bibitem{doi:10.1063/1.4960684}
M.~M. Basko, On the maximum conversion efficiency into the 13.5-nm extreme
  ultraviolet emission under a steady-state laser ablation of tin microspheres,
  Physics of Plasmas 23~(8) (2016) 083114.
\newblock \href {https://doi.org/10.1063/1.4960684}
  {\path{doi:10.1063/1.4960684}}.

\bibitem{stewart_lowering_1966}
J.~C. Stewart, J.~{Pyatt, Kedar D.},
  \href{http://adsabs.harvard.edu/doi/10.1086/148714}{{Lowering of Ionization
  Potentials in Plasmas}}, The Astrophysical Journal 144 (1966) 1203.
\newblock \href {https://doi.org/10.1086/148714} {\path{doi:10.1086/148714}}.
\newline\urlprefix\url{http://adsabs.harvard.edu/doi/10.1086/148714}

\bibitem{doi:10.1063/1.1724509}
G.~Ecker, W.~Kr{\"o}ll,
  \href{https://aip.scitation.org/doi/abs/10.1063/1.1724509}{Lowering of the
  ionization energy for a plasma in thermodynamic equilibrium}, The Physics of
  Fluids 6~(1) (1963) 62--69.
\newblock \href
  {http://arxiv.org/abs/https://aip.scitation.org/doi/pdf/10.1063/1.1724509}
  {\path{arXiv:https://aip.scitation.org/doi/pdf/10.1063/1.1724509}}, \href
  {https://doi.org/10.1063/1.1724509} {\path{doi:10.1063/1.1724509}}.
\newline\urlprefix\url{https://aip.scitation.org/doi/abs/10.1063/1.1724509}

\bibitem{Rozsnyai1977a}
B.~F. Rozsnyai,
  \href{http://www.sciencedirect.com/science/article/B6TVR-46FVDVY-17/2/fc5aa480d46b12213ba118b0f7049479}{{Spectral
  lines in hot dense matter}}, Journal of Quantitative Spectroscopy and
  Radiative Transfer 17~(1) (1977) 77--88.
\newblock \href {https://doi.org/10.1016/0022-4073(77)90142-X}
  {\path{doi:10.1016/0022-4073(77)90142-X}}.
\newline\urlprefix\url{http://www.sciencedirect.com/science/article/B6TVR-46FVDVY-17/2/fc5aa480d46b12213ba118b0f7049479}

\bibitem{NLTE_site}
{The Non-LTE Code Comparison Workshop},
  \href{http://nlte.nist.gov/}{http://nlte.nist.gov/}.
\newline\urlprefix\url{http://nlte.nist.gov/}

\bibitem{RPHDM18}
\href{https://indico.desy.de/indico/event/18869/overview}{{The 2018
  International Workshop on Radiative Properties of Hot Dense Matter}} (2018).
\newline\urlprefix\url{https://indico.desy.de/indico/event/18869/overview}

\bibitem{Voropinov1970}
A.~Voropinov, G.~Gandel'man, V.~Podval'nyi, {ELECTRONIC ENERGY SPECTRA AND THE
  EQUATION OF STATE OF SOLIDS AT HIGH PRESSURES AND TEMPERATURES}, Soviet
  Physics Uspekhi 13~(1) (1970) 56--72.
\newblock \href {https://doi.org/10.1070/PU1970v013n01ABEH004198}
  {\path{doi:10.1070/PU1970v013n01ABEH004198}}.

\bibitem{Cowan1981}
R.~D. Cowan, {The theory of atomic structure and spectra}, University of
  California Press, 1981.

\bibitem{Gu2008}
M.~F. Gu, The flexible atomic code, Can. J. Phys. 86~(5) (2008) 675--689.
\newblock \href {https://doi.org/10.1139/P07-197} {\path{doi:10.1139/P07-197}}.

\bibitem{Novikov2009}
V.~G. Novikov, K.~N. Koshelev, A.~D. Solomyannaya, Radiative unresolved spectra
  atomic model, in: 16th International Conference on Atomic Processes in
  Plasmas, Monterey, CA, 2009.

\bibitem{PhysRevLett.109.065002}
O.~Ciricosta, S.~M. Vinko, H.-K. Chung, B.-I. Cho, C.~R.~D. Brown, T.~Burian,
  J.~Chalupsk\'y, K.~Engelhorn, R.~W. Falcone, C.~Graves, V.~H\'ajkov\'a,
  A.~Higginbotham, L.~Juha, J.~Krzywinski, H.~J. Lee, M.~Messerschmidt, C.~D.
  Murphy, Y.~Ping, D.~S. Rackstraw, A.~Scherz, W.~Schlotter, S.~Toleikis, J.~J.
  Turner, L.~Vysin, T.~Wang, B.~Wu, U.~Zastrau, D.~Zhu, R.~W. Lee, P.~Heimann,
  B.~Nagler, J.~S. Wark,
  \href{https://link.aps.org/doi/10.1103/PhysRevLett.109.065002}{Direct
  measurements of the ionization potential depression in a dense plasma}, Phys.
  Rev. Lett. 109 (2012) 065002.
\newblock \href {https://doi.org/10.1103/PhysRevLett.109.065002}
  {\path{doi:10.1103/PhysRevLett.109.065002}}.
\newline\urlprefix\url{https://link.aps.org/doi/10.1103/PhysRevLett.109.065002}

\bibitem{IGLESIAS2013103}
C.~A. Iglesias, P.~A. Sterne,
  \href{http://www.sciencedirect.com/science/article/pii/S1574181812001334}{Fluctuations
  and the ionization potential in dense plasmas}, High Energy Density Physics
  9~(1) (2013) 103 -- 107.
\newblock \href {https://doi.org/https://doi.org/10.1016/j.hedp.2012.11.007}
  {\path{doi:https://doi.org/10.1016/j.hedp.2012.11.007}}.
\newline\urlprefix\url{http://www.sciencedirect.com/science/article/pii/S1574181812001334}

\bibitem{NovikovRalchenkoBook}
V.~G. Novikov,
  \href{http://link.springer.com/10.1007/978-3-319-27514-7{\_}5}{{Average Atom
  Approximation in Non-LTE Level Kinetics}}, in: Y.~Ralchenko (Ed.), Modern
  Methods in Collisional-Radiative Modeling of Plasmas, springer s Edition,
  Springer International Publishing, Switzerland, 2016, Ch.~5, pp. 105--126.
\newblock \href {https://doi.org/10.1007/978-3-319-27514-7_5}
  {\path{doi:10.1007/978-3-319-27514-7_5}}.
\newline\urlprefix\url{http://link.springer.com/10.1007/978-3-319-27514-7{\_}5}

\end{thebibliography}

\end{document}